\documentclass[fontsize=12pt]{article}
\usepackage[natbibapa]{apacite}
\makeatletter
\NAT@longnamesfalse
\makeatother
\AtBeginDocument{%
}

\usepackage{amsmath,amsthm}
\usepackage{xcolor}
\usepackage{enumerate}
\usepackage{graphicx} 
\usepackage{mathtools}
\usepackage{caption,subcaption}
\usepackage{authblk}
\usepackage{multirow}
\usepackage{tablefootnote}
\usepackage{comment}
\usepackage{algorithm,algpseudocode}
\usepackage{url}

\newtheorem{theorem}{Theorem}
\newtheorem{lemma}[theorem]{Lemma}

\usepackage{setspace}

\title{\bf\Large Analyzing Zero-Truncated Recurrent Events by Stratified Regression with Time-Varying Coefficients}

\author[1]{\bf\normalsize Anqi A. Chen \thanks{email: aca142@sfu.ca}}
\author[1]{\bf X. Joan Hu \thanks{email: joanh@stat.sfu.ca}}
\author[2]{\bf Rhonda J. Rosychuk \thanks{email: rhonda.rosychuk@ualberta.ca}}
\author[3]{\bf Leilei Zeng \thanks{email: lzeng@uwaterloo.ca}}
\affil[1]{\normalsize Department of Statistics and Actuarial Science, Simon Fraser University, Burnaby, British Columbia, Canada}
\affil[2]{Department of Pediatrics, University of Alberta, \newline Edmonton, Alberta, Canada}
\affil[3]{Department of Statistics and
Actuarial Science, University of Waterloo, Waterloo, Ontario, Canada}

\usepackage[margin=2in]{geometry}
\begin{document}



\date{} 
\maketitle

\begin{abstract}
This paper presents a strategy for analyzing zero-truncated recurrent events data. Motivated by a pediatric mental health care (PMHC) program, we are particularly concerned with how the event occurrence depends on the occurrences in the past. We consider a stratified Cox regression model with time-varying coefficients and propose a procedure for estimating the model parameters using the zero-truncated data integrated with population census information.  We evaluate the finite-sample performance of the proposed estimator through simulation and establish its asymptotic properties. Data from the PMHC program are used throughout the paper to motivate and to illustrate the proposed approach.

\end{abstract}


\noindent{\it keywords:}
Conditional intensity function;
Cox regression model;
Doubly censored recurrent event data; 
Partially known stratification;
Supplementary information.


\section{Introduction}\label{sec:Introduction}

This study is motivated by an ongoing pediatric mental health care program that focuses on mental health-related emergency department (MHED) visits made by Alberta, Canada residents under 18 years old. The MHED data are extracted from population-based administrative databases that provide large data but do not capture all the information we need. 
Specifically, we are interested in making inferences about Alberta residents under age 18, including eligible individuals who do not have MHED visits. For those without emergency department (ED) presentations, covariates and observation periods by age are missing. \cite{Hu&Lawless1966_supInfo} developed a method for analyzing truncated lifetime data with some supplementary information. They considered two approaches for obtaining additional information on covariate and observation times (or censoring times in their context): drawing a random sample from subjects without failures or directly using readily available population information. They noted that the sample in the second approach was not necessarily independent of the primary event time data and could be provided in an aggregate form. 
We adapt their second approach in this paper.

The available data include all ED records of Alberta residents under 18 years old from April 1, 2010 to March 31, 2017 and demographic and geographic information of the individuals with MHED visits. We formulate the MHED visit records as recurrent events with the observations subject to zero-truncation. That is, the study subjects and their observation times are only observed if the individuals experienced at least one event over the data extraction window. \cite{Hu&Lawless1966_supinf_rateMean} proposed nonparametric methods to estimate rate and mean functions for zero-truncated recurrent events from a Poisson or arbitrary process when the population size and distribution of observation times were at least approximately known. Using similar MHED data, \cite*{Yi_Hu_Rosychuk2020} compared the event patterns across two decades under an extended Cox regression model with time-varying coefficients. They proposed an estimation procedure based on zero-truncated recurrent event data combined with population census information.

Our research aims to study the event patterns and to understand how the event occurrence depends on the prior occurrences. The second aim leads us to consider an intensity-based model, in contrast to marginal models like those developed by \cite{rate_PEPEMS1993SGDA}, \cite{rate_COOKRICHARDJ.1997MAOR}, \cite{rate_LinD.Y.2000Srft}, \cite{marginal_LiShanshan2016Reda},
 \cite{rate_HuangMing-Yueh2023Iseo}, and \cite*{marginal-Xu2024}. Intensity-based models fully specify the counting processes and depend on the event history, whereas marginal models partially specify the processes and do not condition on the history \citep{Cook.Lawless.2007}. In our study, the history is partially known since we only observe the full history for individuals born during the study period (that is, the period during which MHED visits occurred). This feature calls for a modeling approach that lies between these two. \cite*{Chen&Hu&Rosychuk2025} proposed an intensity-based model with time-varying stratification to fully specify the dependence on the history and developed a method for estimating time-independent coefficients when the stratification is only partially known. Time-varying stratification is also used in Cox-type models by, for example,  \cite{startificat_ChangShu-Hui1999}, \cite{hu2011analysis}, \cite*{PWP1981}, \cite{stratificat_Wang2014}, and \cite{stratification_ZhongYujie2021}; they defined strata based on the number of previous events. 

 Different from \cite*{Chen&Hu&Rosychuk2025} focusing on constant coefficients, this study is concerned with the estimation of time-dependent regression coefficients using the zero-truncated data. There is rich literature on estimating time-varying coefficients in other contexts. \cite{Tv_coef_CAIZONGWU2003} and \cite*{TV_ceof_TianLu2005} presented local linear partial likelihood methods under a Cox regression model \citep*{Cox_reg_model}. \cite{Hu_Rosychuk2016} adapted Cai and Sun's method to investigate doubly censored (both left- and right-censored) recurrent event data. More recent studies on the local estimation include those by \cite*{TV_coef_Wu2021} and \cite{Yi_Hu_Rosychuk2020}. Alternatively, spline-based approaches have been proposed by \cite{TV_coef_AmorimLeilaD.2008}, \cite{TV_coef_HeKevin2022}, \cite{TV_coef_WuWenbo2022}, and \cite*{TV_coef_DaZhao2025}.
 
The remainder of the paper is organized as follows: Section~\ref{sec:notation_model} introduces the notation and modeling.
Section~\ref{sec:estimation} presents the estimation procedures using zero-truncated recurrent event data integrated with supplementary information. Section~\ref{sec:real_data} shows the application of the proposed approach to the MHED data. Section~\ref{sec:simulation} reports a simulation study conducted to evaluate the numerical performance of the approach. Section~\ref{sec:conclusion} provides final remarks and discussions. All the simulated data in the paper were generated using R (Version 3.6.1). The proposed method was implemented in C++ and executed through the R packages \textit{Rcpp} \citep{Rccp} and \textit{RcppArmadillo} \citep{RcppArmadillo}.

\section{Notation and Modeling}\label{sec:notation_model}
Consider that the target population $\mathcal{P}$ includes all independent individuals under age $A^\star$, where $A^\star=18$ in the MHED study. Let $\mathcal{P}=\mathcal{P}_1\cup \mathcal{P}_0$, where $P_1$ represents a subpopulation having at least one event, and $P_0$ is a subpopulation without events before age $A^\star$. 

We use the personal time (that is, age in years) in this paper. Let $N_i(a)$ represent the number of MHED visits made by subject $i$ since birth up to age $a$ for $i\in \mathcal{P}$ and $a\geq 0$. By convention, $N_i(0)\equiv 0$. Denote the event history of subject $i$ before age $a$ by $\mathcal{H}_i(a)=\sigma\big\{N_i(u): 0\leq u<a\big\}$, the $\sigma$-algebra generated by the event information before age $a$. Define a stratification variable $S_i(a)=S\big\{\mathcal{H}_i(a)\big\}$, and suppose $S_i(a)$ takes a finite number of values and is a left-continuous and non-decreasing function. Let $Z_i$ be the time-independent covariates for subject $i$. We consider the following extended Cox regression model: for $a>0$, $i\in \mathcal{P}$, and $S_i(a)=s$,
\begin{equation}
\lambda(a\mid \mathcal{H}_i(a), Z_i)
=\lambda_{0s}(a) \exp\{\beta_s(a)^{'} Z_i\}.
\label{eq:model}
\end{equation}
This model includes one type of the models in \cite{PWP1981}, the Andersen-Gill model (\citeyear{AGmodel}), and the \citeauthor{Hu_Rosychuk2016}'s model (\citeyear{Hu_Rosychuk2016}) as special cases. More relevant special cases are provided in Table \ref{Tab:submodels}, which are considered in the subsequent numerical analyses for comparison purposes. Importantly, Model (\ref{eq:model}) depends on the event history only through the stratification variable $S_i(a)$, meaning that we only need a summary of the history rather than detailed history information to fully specify the counting process $N_i(\cdot)$. One example of the stratification is, for $a\in(0,A^*)$,
\begin{align}
    S_i(a)=\begin{cases}1 & N_i(a-) = 0\\2 &N_i(a-) > 0\end{cases}.
\label{eq:stratification_variable}
\end{align}
This variable indicates that subject $i$ belongs to stratum $1$ at or before their first event time and to stratum $2$ afterward. Another example is $S_i(a)=N_i(a-)+1$, suggesting that subject $i$ is initially in stratum $1$ and transitions to stratum $k$ after their $(k-1)$-th event time for $k=2,\cdots,N_i(A^\star-)+1$. Allowing for different stratification rules enhances the flexibility of our proposed model.

\begin{table}[ht!]
\centering
\caption{Relevant model specifications}
\label{Tab:submodels}
\small
\vspace{-0.25cm}
\begin{tabular}{lcc}
\hline
\hline
$\text{Model}^\star$ & Baseline & Coefficients\\
  \hline
\multicolumn{3}{l}{Time-independent coefficients}\\
  \hline
  $\text{(NNC)}^\dag$ & $\lambda_{0s}(a)=\lambda_{0}(a)$ & $\beta_s(a)=\beta$\\ 
(SNC) & $\lambda_{0s}(a)=\lambda_{0s}(a)$ & $\beta_s(a)=\beta$\\ 
(NSC) & $\lambda_{0s}(a)=\lambda_{0}(a)$ & $\beta_s(a)=\beta_s$\\ 
(SSC) & $\lambda_{0s}(a)=\lambda_{0s}(a)$ & $\beta_s(a)=\beta_s$\\
  \hline
\multicolumn{3}{l}{Time-varying coefficients}\\
  \hline
  $\text{(NNV)}^{\ddag}$ & $\lambda_{0s}(a)=\lambda_{0}(a)$ & $\beta_s(a)=\beta(a)$\\ 
(SNV) & $\lambda_{0s}(a)=\lambda_{0s}(a)$ & $\beta_s(a)=\beta(a)$\\ 
(NSV) & $\lambda_{0s}(a)=\lambda_{0}(a)$ & $\beta_s(a)=\beta_s(a)$\\ 
$\text{(SSV)}^\S$ & $\lambda_{0s}(a)=\lambda_{0s}(a)$ & $\beta_s(a)=\beta_s(a)$\\
   \hline
   \hline
\end{tabular}
\begin{flushleft}
    \scriptsize{\noindent$\star$: The first two letters indicate whether stratification is applied to the baseline and coefficients, respectively, with N for no stratification and S for stratification. The last letter specifies whether the coefficients are constant (C) or time-varying (V);\\
    $\dag$: Model (NNC) is the Andersen-Gill model;\\
    $\ddag$: Model (NNV) is the Hu and Rosychuk's model;\\
    \vspace{-0.1cm}
    $\S$: Model (SSV) is Model (1).}
\end{flushleft}
\end{table}

Model (\ref{eq:model}) has an alternative representation. Let $X_i(a)$ be a column vector of dummy variables corresponding to the categorical variable $S_i(a)$ at a fixed $a$. Model (\ref{eq:model}) can be then rewritten as 
\small
\begin{equation*}
\lambda(a\mid \mathcal{H}_i(a), Z_i)
=\lambda_{0}(a) \exp\{\alpha(a)^{'}X_i(a)+\beta(a)^{'}Z_i+ \kappa(a)^{'}(X_i(a)\otimes Z_i)\}.
\end{equation*}
\normalsize
Here, $X_i(a)$ and $Z_i$ are treated as column matrices and $\otimes$ denotes the Kronecker product. This alternative formulation includes the model given by \cite{Cook.Lawless.2007} as a special case with $x_i(t)=x_i$: 
\begin{equation*}
\lambda_i(t\mid \mathcal{H}_i(t))
=\lambda_{0}(t) \exp\{x_i(t)^{'}\beta+\gamma N_i^\star(t^-)\},
\end{equation*}
where $N_i^\star(t)=N_i(t)$ for $N_i(t)\leq 5$ and $N_i^\star(t)=5$ for $N_i(t)>5$. 


Let $[W_L, W_R]$ denote a predetermined data extraction window in the calendar time. The sample $\mathcal{O}$ consists of all residents under age $A^\star$ in a certain region during $[W_L, W_R]$. For example, $\mathcal{O}$ includes all Alberta residents under 18 years old during April 1, 2010 to March 31, 2017 in our MHED study. The sample $\mathcal{O}$ contains two subsets $\mathcal{O}_1$ and $\mathcal{O}_0$, such that $\mathcal{O}=\mathcal{O}_1\cup \mathcal{O}_0$. Individuals in $\mathcal{O}_1$ experience at least one event, corresponding to the MHED cohort in the study, while those in $\mathcal{O}_0$ have no events within $[W_L, W_R]$. We assume that $\mathcal{O}$ is a random sample of the target population $\mathcal{P}$, with $\mathcal{O}_1$ as a random sample of $\mathcal{P}_1$ and $\mathcal{O}_0$ as a random sample of $\mathcal{P}_0$. 
In the MHED study, the administrative databases provide information only for subjects in $\mathcal{O}_1$. To make inferences on $\mathcal{P}$, we need additional information on $\mathcal{O}_0$.  

Let $B_i$ be the birthdate of subject $i$ in the calendar time for $i\in\mathcal{O}_1$. Denote the subject-specific window in the personal time by $(C_{L_i}, C_{R_i}]$, where $C_{L_i}=\max(0, W_L-B_i)$ and
$C_{R_i}=\min(A^\star,W_R-B_i)$. Events occurring outside $(C_{L_i}, C_{R_i}]$ are not observed. 
We assume that $W_L$ and $W_R$ are independent of the events
and $B_i$ is independent of the counting process $N_i(\cdot)$. Define the number of observed events as $N^{\star}_i =N_i(C_{R_i})-N_i(C_{L_i})$ with $N^{\star}_i>0$ for all $i \in \mathcal{O}_1$. We know $N^{\star}_i=0$ for all $i\in\mathcal{O}_0$ even though their birthdates and observation windows are missing.
Let $T_{ij}$ represent the calendar time of the $j$th
observed event made by subject $i$ for $i\in\mathcal{O}_1$ and $j=1,
\cdots,N^{\star}_i$. The event time in age is given by $A_{ij}=T_{ij}-B_i$. The available data are summarized as $\mathcal{Q}_1=\bigcup_{i\in\mathcal{O}_1} \mathcal{Q}_{1i}=\bigcup_{i\in\mathcal{O}_1}\big\{\{dN_i(a):C_{L_i}<a\leq C_{R_i}\}\bigcup_{}^{}\{Z_i\}\big\}$, where $dN_i(\cdot)$ denotes the increments in $N_i(\cdot)$. We assume that there is no tie. In this case, $dN_i(\cdot)=1$ at the observed event times in age and $dN_i(\cdot)=0$ otherwise. 

\section{Estimation Procedure}\label{sec:estimation}

The statistical goal is to estimate the baseline intensity functions $\boldsymbol{\lambda}_{0}(\cdot)=\{\lambda_{0s}(\cdot):s\in\mathcal{S}\}$ and the regression coefficients $\boldsymbol{\beta}(\cdot)=\{\beta_s(\cdot):s\in\mathcal{S}\}$, where $\mathcal{S}$ denotes all the possible values of the stratification variable $S(a)$ for $a>0$. In the MHED study, the full event history is observed only for subjects born during the study period. Therefore, the stratification, defined as a function of the history, is only partially known. In this section, we first present an estimation procedure under an idealized setting where the stratification is fully known. We then turn to the more realistic case where the stratification is only partially known.
 
\bigskip

\noindent3.1\textit{\hspace{0.4cm}Estimation with Fully Known Stratification for Subjects in $\mathcal{O}_1$ Only} \label{sec:estimation_known_S} 

\noindent The available data $\mathcal{Q}_1$ contain information only for subjects in $\mathcal{O}_1$, leaving covariates and observation times missing for subjects in $\mathcal{O}_0$. \cite{Yi_Hu_Rosychuk2020} addressed this issue by incorporating population census data as the supplementary information for $\mathcal{O}$, which includes $\mathcal{O}_0$. 
Extending their method, we use aggregate census data to approximate the missing parts of our estimating functions shown later. The aggregate data provide yearly recorded subgroup counts, such as the number of 4-year-old girls living in Edmonton in 2017. 

We adapt the local linear estimation approach in \cite{Hu_Rosychuk2016} to estimate the time-varying coefficients $\boldsymbol{\beta}(\cdot)$ and further simplify it using a local constant estimation. Specifically, we select constants $\tau_L$ and $\tau_R$ satisfying three conditions: $0<\tau_L<\tau_R<A^\star$, $P(C_{L}<\tau_L)>0$, and $P(C_{R}>\tau_R)>0$. The constants help avoid boundary problems in the local estimation. For a fixed $a\in[\tau_L,\tau_R]$, we can approximate $\beta_s(u)$ with $\beta_s(a)$ for $u$ in a neighbourhood of $a$ and $s\in\mathcal{S}$.

Let $\gamma_s=\beta_s(a)$ and $\boldsymbol{\gamma}=\{\gamma_s:s\in\mathcal{S}\}$. We use a kernel function with a bandwidth $h$, denoted by $K_h(\cdot)=K(\cdot/h)/h$, to assign weights according to the distance between two age points. Define the indicators $Y_i^{(c)}(u)= I\big(u \in (C_{L_i}, C_{R_i}]\big)$ and $Y_i^{(s)}(u)=I\big(u: S_i(u)=s\big)$. We consider the following estimating function of $\gamma_s$, for a fixed $a\in[\tau_L,\tau_R]$:
\begin{align}
U_{s}(\boldsymbol{\gamma};a|\boldsymbol{\lambda}_{0}(\cdot))=\sum_{i\in \mathcal{O}_1}^{}\int_{0}^{A^\star} K_h(u-a) Y_i^{(s)}(u)\big\{Z_i-\tilde{\bar{Z}}_s(\boldsymbol{\gamma};u)\big\} Y_i^{(c)}(u) dN_i(u), 
    \label{eq:EF_coef_S_known}
\end{align}
where $\tilde{\bar{Z}}_s(\boldsymbol{\gamma};u)=\sum_{i\in\mathcal{O}}^{}\hat{E}\big[Y_i^{(s)}(u)$ $Y_i^{(c)}(u)Z_i\exp\{\gamma_s^{'}Z_i\}\big]/\sum_{i\in\mathcal{O}}^{}\hat{E}\big[Y_i^{(s)}(u)$ $Y_i^{(c)}(u)\exp\{\gamma_s^{'}Z_i\}\big]$. The term $\tilde{\bar{Z}}_s(\boldsymbol{\gamma};u)$ approximates $\bar{Z}_s(\boldsymbol{\gamma};u)=\sum_{i\in\mathcal{O}}^{} $ $Y_i^{(s)}(u)Y_i^{(c)}(u)Z_i\exp\{\gamma_s^{'}Z_i\}/\sum_{i\in\mathcal{O}}^{} Y_i^{(s)}(u)Y_i^{(c)}(u)\exp\{\gamma_s^{'}Z_i\}$ using census data. We assume that $\{N_i(\cdot),Z_i,B_i\}$ for $i=1,\cdots,|\mathcal{O}|$ are independent and identically distributed. Then, with $q=0,1,2$,
\small
    \begin{align*}
        \begin{split}
        &\sum_{i\in\mathcal{O}}^{}E\big[Y_i^{(s)}(u)Y_i^{(c)}(u)Z_i^{\otimes q}\exp\{\gamma_s^{'}Z_i\}\big]\\
        &\hspace{0.3cm}=\sum_{z\in \mathcal{Z}}^{}\Bigl\{
P\bigl(Y_1^{(s)}(u)=1|Z_1=z\bigr)z^{\otimes q}\exp\{\gamma_s^{'}z\}|\mathcal{O}|P\bigl(Y_1^{(c)}(u)=1,Z_1=z\bigr)\Bigr\},\\
        \end{split}
    \end{align*}
\normalsize
 where $\mathcal{Z}$ denotes all the possible covariate combinations. Here, $b^{\otimes 0}=1$, $b^{\otimes 1}=b$, and $b^{\otimes 2}=bb^{'}$. We consider that all the covariates are discrete and take a finite number of values. We can use the aggregate census data to estimate $|\mathcal{O}|P\bigl(Y_1^{(c)}(u)=1,Z_1=z\bigr)$; that is, $|\mathcal{O}|\hat{P}\bigl(Y_1^{(c)}(u)=1,Z_1=z\bigr)=\sum_{l}\mathcal{C}(l,z,\lfloor u \rfloor)$. The $\mathcal{C}(l,z,\lfloor u \rfloor)$ represents the count of individuals at age $\lfloor u \rfloor$ with covariates $Z=z$ in the calendar year $l$. 
Then, $\sum_{i\in\mathcal{O}}^{}\hat{E}\big[Y_i^{(s)}(u)Y_i^{(c)}(u)Z_i^{\otimes q}\exp\{\gamma_s^{'}Z_i\}\big]$ can be rewritten as $\sum_{z\in \mathcal{Z}}^{}\Bigl\{
P\bigl(Y^{(s)}$ $(u)$ $=1|Z=z\bigr)z^{\otimes q}\exp\{\gamma_s^{'}z\}\bigl[\sum_{l}\mathcal{C}(l,z,$ $\lfloor u \rfloor)\bigr]\Bigr\}$. If the census data are recorded monthly or daily, we can get a better approximation of $\bar{Z}_s(\boldsymbol{\gamma};u)$.

Define $\Lambda_{0s}(a)=\int_0^{a}\lambda_{0s}(u)du$ for $a\in(0,A^\star) $. Given $\boldsymbol{\beta}(\cdot)$, the estimating function of the cumulative baseline intensity function $\Lambda_{0s}(\cdot)$ is, for $s\in\mathcal{S}$,
\small
\begin{align}
\begin{split}
    V_{s}(\boldsymbol{\Lambda}_{0}(\cdot);&a\big|\boldsymbol{\beta}(\cdot))=\sum_{i\in \mathcal{O}_1}^{}Y_i^{(s)}(a)Y_i^{(c)}(a)dN_i(a)\\
    &-\sum_{z\in \mathcal{Z}}^{}\Bigl\{
P\bigl(Y^{(s)}(a)=1|Z=z\bigr)\exp\{\beta_s(a)^{'}z\}\bigl[{\sum_{l}\mathcal{C}(l,z,\lfloor a \rfloor)}\bigr]\Bigr\}d\Lambda_{0s}(a).
\label{eq:EE_baseline_S_known}
\end{split}
\end{align}
\normalsize
Since the probability $P\bigl(Y^{(s)}(a)=1|Z=z\bigr)$ may involve both the coefficients and baselines (see Appendix \ref{subsec:Append_P_s_Z} for an example), we jointly solve $U_{s}(\boldsymbol{\gamma};a|\boldsymbol{\lambda}_{0}(\cdot))=\mathbf{0}$ and $V_{s}(\boldsymbol{\Lambda}_{0}(\cdot);a\big|\boldsymbol{\beta}(\cdot))=0$ to obtain the local constant estimator $\hat{\beta}_s(a)$ for $a\in[\tau_L,\tau_R]$ and the estimated cumulative baseline intensity function: for $a\in(0,A^\star)$ and $s\in\mathcal{S}$, by convention, $0/0=0$,
\begin{align}
\begin{split}
\label{eq:Lambda_est_S_known}
 \hat{\Lambda}_{0s}(a\big|\boldsymbol{\beta}(\cdot))=\sum_{i\in \mathcal{O}_1}\int_{0}^{a} \frac{Y_i^{(s)}(u)Y_i^{(c)}(u)}{\sum_{i\in\mathcal{O}}^{}\hat{E}\big[Y_i^{(s)}(u)Y_i^{(c)}(u)\exp\{\beta_s(u)^{'}Z_i\}\big]}dN_i(u).
\end{split}
\end{align}
\normalsize
 Additionally, set $\hat{\beta}_s(a)=\hat{\beta}_s(\tau_L)$ for $a\in(0,\tau_L)$, and $\hat{\beta}_s(a)=\hat{\beta}_s(\tau_R)$ for $a\in(\tau_R,A^\star)$. 
Plugging in $\hat{\boldsymbol{\beta}}(\cdot)=\{\hat{\beta}_{s}(\cdot):s\in\mathcal{S}\}$, we obtain the Breslow estimator $\hat{\Lambda}_{0s}(\cdot\big|\hat{\boldsymbol{\beta}}(\cdot))$ \citep{breslow_est1972}. The baseline intensity function can be estimated by differentiating $\hat{\Lambda}_{0s}(a\big|\hat{\boldsymbol{\beta}}(\cdot))$ with respect to $a$ for $a\in(0,A^\star)$ and $s\in\mathcal{S}$. 
\bigskip

\noindent3.2\textit{\hspace{0.4cm}Estimation with Partially Known Stratification} \label{sec:estimation_P_avail_S}

\noindent When the stratification is only partially known, equations (\ref{eq:EF_coef_S_known}) and (\ref{eq:Lambda_est_S_known}) in Section 3.1 are not always evaluable, since the value of $Y^{(s)}_i(a)$ is not always known for $i\in\mathcal{O}_1$. To address this situation, we replace $Y^{(s)}_i(a)$ with its conditional probability $P\bigl(Y^{(s)}_i(a)=1|\mathcal{Q}_{1i}\bigr)$, leading to the following estimating functions: for $s\in\mathcal{S}$ and $a\in[\tau_L,\tau_R]$,
\small
\begin{align}
\begin{split}
\tilde{U}_{s}(\boldsymbol{\gamma};a|\boldsymbol{\lambda}_{0}(\cdot))
&= \sum_{i\in \mathcal{O}_1}^{}\int_{0}^{A^\star} K_h(u-a) P\bigl(Y_i^{(s)}(u)=1|\mathcal{Q}_{1i}\bigr)\big\{Z_i\\
&\hspace{3.5cm}-\tilde{\bar{Z}}_s(\boldsymbol{\gamma};u)\big\} Y_i^{(c)}(u) dN_i(u),
    \label{eq:EF_coef_partial_s}
    \end{split}
    \end{align}
\normalsize
and for $a\in(0,A^\star)$,
\small
\begin{align}
\begin{split} 
 \tilde{\Lambda}_{0s}(a\big|\boldsymbol{\beta}(\cdot))=\sum_{i\in \mathcal{O}_1}\int_{0}^{a} \frac{P\bigl(Y_i^{(s)}(u)=1|\mathcal{Q}_{1i}\bigr)}{\sum_{i\in\mathcal{O}}^{}\hat{E}\big[Y_i^{(s)}(u)Y_i^{(c)}(u)\exp\{\beta_s(u)^{'}Z_i\}\big]}Y_i^{(c)}(u)dN_i(u).
 \end{split}
 \label{eq:baseline_EF_partial_S}
\end{align}
\normalsize
We consider $P(Y_i^{(s)}(u)=1|\mathcal{Q}_{1i})$ to be specified subject to Model (\ref{eq:model}) to carry out the estimation procedure. Appendix \ref{subsec:Append_p_s_data} shows an example of calculating the conditional probability using the stratification variable (\ref{eq:stratification_variable}).

We jointly solve the estimating equations for the regression coefficients and the cumulative baseline intensity functions to obtain the estimators, denoted by $\tilde{\beta}_s(a)$ and $\tilde{\Lambda}_{0s}(a)$, for $s\in\mathcal{S}$ and $a\in(0,A^\star)$. The asymptotic properties of 
$\tilde{\boldsymbol{\beta}}(\cdot)=\{\tilde{\beta}_s(\cdot):s\in\mathcal{S}\}$, which is our main interest, are described as follows.
\bigskip

\noindent \textbf{Proposition 1}. \textit{Under conditions (I)-(VII)} listed in Appendix \ref{sec:Appendix_proofs}, the estimator $\tilde{\boldsymbol{\beta}}(\cdot)$ has the following asymptotic properties: For $a\in[\tau_L,\tau_R]$,

 \begin{enumerate}[(i)]
    \item Pointwise consistency: $\tilde{\boldsymbol{\beta}}(a) \xrightarrow{a.s.} \boldsymbol{\beta}_0(a)$ as $n\rightarrow \infty$, where $\boldsymbol{\beta}_0(a)$ represents the true functions, and $n=|\mathcal{O}|$.
     \item Weak convergence: With the bandwidth $h=O(n^{-v})$ for $1/2<v<1$, $\sqrt{nh}(\tilde{\boldsymbol{\beta}}(a)-\boldsymbol{\beta}_0(a))\xrightarrow{d} N \big(\mathbf{0},AV(\boldsymbol{\beta}_0(a))\big)$ as $n\rightarrow \infty$.
 \end{enumerate}
The asymptotic derivation follows the approaches of \cite{Hu_Rosychuk2016}, \cite{Yi_Hu_Rosychuk2020}, and \cite{Chen&Hu&Rosychuk2025}. An outline of the proof for Proposition 1, together with the sufficient conditions, is presented in Appendix \ref{sec:Appendix_proofs}.

To estimate the asymptotic variance $AV(\boldsymbol{\beta}_0(\cdot))$, we consider a resampling method based on Poisson multipliers with a mean and variance of $1$ \citep*{Poisson_multi} :
\begin{enumerate}
    \item 	Generate a set of multipliers $W^{(b)}=(W_1^{(b)},\cdots,W_{|\mathcal{O}_1 |}^{(b)} )'$, where $W_i^{(b)}\overset{\mathrm{iid}}{\sim}\text{Poisson}(1)$ for $i=1,\cdots, |\mathcal{O}_1 |$. The multipliers are independent of the data.
    \item 	Given a realization of $W^{(b)}$, jointly solve the following equations for $\gamma_s=\beta_s(a)$ and $\Lambda_{0s}(a)$: for $a\in[\tau_L,\tau_R]$,
\scriptsize
\begin{align*}
\begin{split}
    \sum_{i\in \mathcal{O}_1}^{}\int_{0}^{A^\star} &K_h(u-a) P\bigl(Y_i^{(s)}(u)=1|\mathcal{Q}_{1i}\bigr)\big\{Z_i-\tilde{\bar{Z}}_s(\boldsymbol{\gamma};u)\big\} Y_i^{(c)}(u) dN_i(u) {
    W_i^{(b)}}=\mathbf{0},
    \end{split}
\end{align*}
\normalsize
and for $a\in(0,A^\star)$,
\scriptsize
\begin{align*}
\begin{split}
 \tilde{\Lambda}_{0s}^{(b)}(a\big|\boldsymbol{\beta}(\cdot))=\sum_{i\in \mathcal{O}_1}\int_{0}^{a} \frac{P\bigl(Y_i^{(s)}(u)=1|\mathcal{Q}_{1i}\bigr)}{\sum_{i\in\mathcal{O}}^{}\hat{E}\big[Y_i^{(s)}(u)Y_i^{(c)}(u)\exp\{\beta_s(u)^{'}Z_i\}\big]}Y_i^c(u)dN_i(u) W_i^{(b)}.
    \end{split}
\end{align*}
\normalsize
\item Repeat Steps 1 and 2 for $B$ times to obtain the resampling estimates $\widetilde{\boldsymbol{\beta}}^{(1)}(\cdot),\cdots, \widetilde{\boldsymbol{\beta}}^{(B)}(\cdot)$. The estimated asymptotic variance of $\widetilde{\boldsymbol{\beta}}(\cdot)$ is the sample variance of these estimates.
\end{enumerate}
 In Step 1, the multipliers may also be generated from a standard normal distribution \citep*{Lin1993_normal_multi}. The key is to generate multipliers from a distribution with a variance of 1.

In practice, it is hard to jointly solve the estimating equations of coefficients and (cumulative) baselines. Thus, we propose Algorithm \ref{alg:approch_1} to estimate $\boldsymbol{\lambda}_{0}(\cdot)$ and $\boldsymbol{\beta}(\cdot)$ iteratively.

\begin{algorithm}[H]
\caption{Estimate $\boldsymbol{\lambda}_{0}(\cdot)$ and $\boldsymbol{\beta}(\cdot)$}\label{alg:approch_1}
\begin{algorithmic}
\State Let $\boldsymbol{\lambda}_{0}^{(r)}(\cdot)$ and $\boldsymbol{\beta}^{(r)}(\cdot)$ represent the estimates of the baseline intensity functions and regression coefficients in the $r$-th iteration, for $r=0,1,\cdots$. The $\boldsymbol{\lambda}_{0}^{(0)}(\cdot)$ and $\boldsymbol{\beta}^{(0)}(\cdot)$ are predetermined initial functions.
\begin{enumerate}
    \item Given $\boldsymbol{\lambda}_{0}^{(r)}(\cdot)$ and $\boldsymbol{\beta}^{(r)}(\cdot)$, estimate $P\bigl(Y_i^{(s)}(a)=1|\mathcal{Q}_{1i}\bigr)$ for $i\in \mathcal{O}_1$ and $P\bigl(Y^{(s)}(a)=1|Z=z\bigr)$ for $z\in\mathcal{Z}$. With $a\in[\tau_L,\tau_R]$, solve $\tilde{U}_{s}(\boldsymbol{\gamma};a|\boldsymbol{\lambda}_{0}(\cdot))$ $=\mathbf{0}$ for $\gamma_s=\beta_s(a)$ to further obtain $\boldsymbol{\beta}^{(r+1)}(\cdot)$.
    \item  With $\boldsymbol{\beta}^{(r+1)}(\cdot)$ and the estimated probabilities, obtain $\boldsymbol{\lambda}_{0}^{(r+1)}(\cdot)$ based on $\tilde{\Lambda}_{0s}(\cdot\big|\boldsymbol{\beta}^{(r+1)}(\cdot))$.
\end{enumerate}
Repeat Steps 1 and 2 until $||\beta_s^{(r)}(a)-\beta_s^{(r-1)}(a)||_1\big/||\beta_s^{(r-1)}(a)||_1\leq \tau^\star$ for $\forall a\in[\tau_L,\tau_R]$ and $\forall s \in \mathcal{S}$, where $\tau^\star$ is a predetermined tolerance level. 
\end{algorithmic}
\end{algorithm}

\noindent The sequence limits in the above algorithm are the estimated coefficients and baseline intensity functions derived using our proposed approach.
\bigskip

\section{Analysis of the MHED Dataset}\label{sec:real_data}
The MHED data include the ED records of Alberta residents under 18 years old from April 1, 2010 to March 31, 2017 and geographic and demographic information on the MHED cohort. The MHED records were extracted from the National Ambulatory Care Reporting System (\citealp{NACRS}) developed by the Canadian Institute for Health Information, and the demographic and geographic data were obtained from a linked annual Cumulative Registry File. 

We considered two covariates: \texttt{sex} (male vs. female) and \texttt{region of residence} (Edmonton, Calgary vs. all other regions [\texttt{Others}]), with females and other regions set as the reference levels. Since these covariates were relatively stable, we used information from each subject’s first observed visit and treated them as time-independent variables in the subsequent analyses.

\begin{table*}[ht!]
\centering
\caption{Summary of the recorded MHED visits (2010-2017)  }
\small
\tabcolsep=2pt
\label{Tab:visit_summary}
\begin{tabular*}{\columnwidth}{@{\extracolsep\fill}c|r|rrr|rr|rrr@{\extracolsep\fill}}
\hline
\multirow{2}{*}{Visit} & \multirow{2}{*}{Total} & \multicolumn{3}{c|}{Age in years}  & \multicolumn{2}{c|}{Sex}  & \multicolumn{3}{c}{Region of residence}                                                           \\ \cline{3-10} 

 &  & 0-5  & \multicolumn{1}{c}{6-11} & \multicolumn{1}{c|}{12-17} & Female        & \multicolumn{1}{c|}{Male} & \multicolumn{1}{r}{Edmonton} & \multicolumn{1}{r}{Calgary}  & \multicolumn{1}{r}{Others}\\ \hline
 
1st  & 33299  & \multicolumn{1}{c}{1391} & \multicolumn{1}{c}{3775}               & \multicolumn{1}{c|}{28133}  & \multicolumn{1}{r}{19210} & 14089                    & 8365   & 11579   & \multicolumn{1}{r}{13355} \\ 

2nd & 10971  & \multicolumn{1}{r}{60} & 744  & 10167  & \multicolumn{1}{r}{7011} & 3960                   
 & 2822  & 3815 & \multicolumn{1}{r}{4334} \\ 

3rd  & 5243  & 5 & 272& 4966  & \multicolumn{1}{r}{3508} & 1735                    & 1372   & 1855 & \multicolumn{1}{r}{2016} \\ 

$\geq 4$th & 8653  & \multicolumn{1}{r}{0} & 298  & 8355  & \multicolumn{1}{r}{6208} & 2445  
 & 2576  & 2961 & \multicolumn{1}{r}{3116} \\ \hline
Total & 58166  &1456 &5089 &51621                 
&35937 & 22229                  
 & 15135 & 20210  & 22821\\ \hline
\end{tabular*}
\end{table*}

\begin{table}[ht!]
\centering
\caption{Population census information of Alberta residents under 18 years old in 2013}
\label{Tab:census2013}
\small

\begin{tabular}{r|rrr|r}
  \hline
Sex  & Edmonton & Calgary  & Others &Total\\ 
  \hline
Female  &128545  &157901   &141594  &428040\\ 
  Male &135368  &166656   &149149   &451173\\ 
    \hline
  Total  &263913  &324557   &290743 & 879213\\
   \hline
\end{tabular}
\end{table}

From 2010 to 2017, a total of 33,299 Alberta children and youth made 58,166 MHED visits. About 67.1\% of subjects had only one MHED visit, and 17.2\% made two visits. The majority of the MHED visits (88.7\%) were made by individuals over 11 years old. Table \ref{Tab:visit_summary} provides a summary of the MHED visits in the dataset. Since covariate values were taken from the first observed events, the first row of Table 2 offers a brief summary of the MHED cohort. In the cohort, the majority of subjects were girls, and about 40\% resided outside the Edmonton and Calgary regions. However, there were more boys and Calgary residents in the Alberta population under 18 years old according to the census data (see Table \ref{Tab:census2013}). The differences further indicate that the MHED cohort is not representative of the Alberta population under 18 years old.

We conducted analyses of the MHED data under all the models listed in Table \ref{Tab:submodels}; however, only the results from Models (SSV) and (SSC) are presented in this section. \cite{Chen&Hu&Rosychuk2025} examined both constant and arbitrary baselines for the same dataset. The estimated arbitrary baselines were significantly different from straight lines, indicating that the assumption of constant baselines was not appropriate. Therefore, we did not consider constant baselines in the real data analysis. In the implementation, we discretized age using two months as one age unit. The Epanechnikov kernel, $K_h(v)=3/(4h)\big\{1-(v/h)^2\big\}$ for $|v|\leq1$, was applied with the bandwidth $h=9$ units (equivalent to 1.5 years). We set $\tau_L=9$ units (1.5 years old) and $\tau_R=105$ units (17.5 years old). We considered the stratification variable (\ref{eq:stratification_variable}) introduced in the previous example.
That is, subject $i$ is initially in stratum 1 and moves to stratum 2 after their first event time. We chose it because most subjects had only one or two visits, and those who experienced the first MHED visit were generally at higher risk for subsequent visits. The first visit might trigger further events; for instance, subjects might return to the EDs for additional care if other mental health services were not accessed. 

\begin{figure}[t!] 
\begin{subfigure}{0.48\textwidth}
\includegraphics[width=\linewidth]{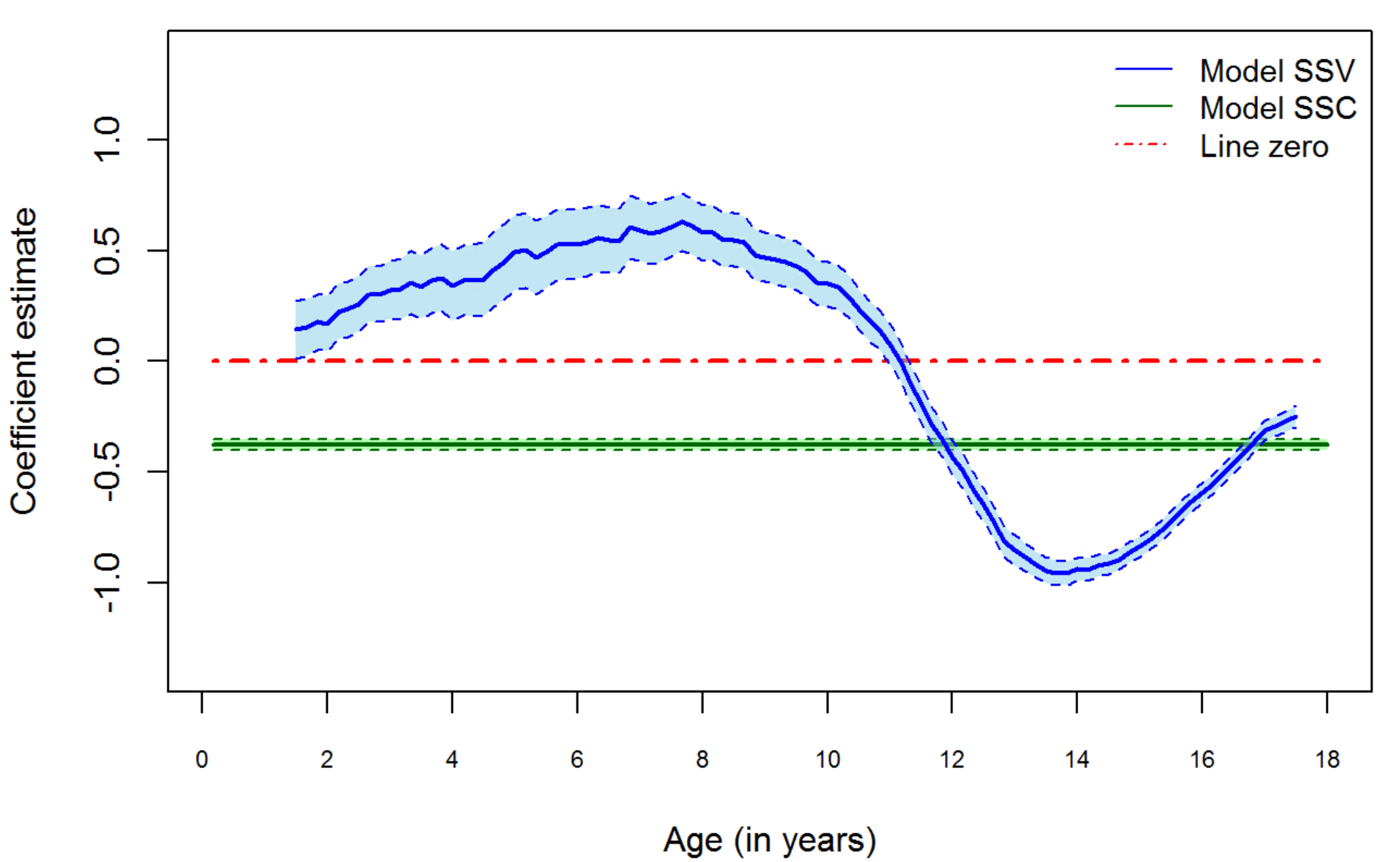}
\caption{Sex (male vs. female); stratum 1} \label{fig:R_a}
\end{subfigure}\hspace*{\fill}
\begin{subfigure}{0.48\textwidth}
\includegraphics[width=\linewidth]{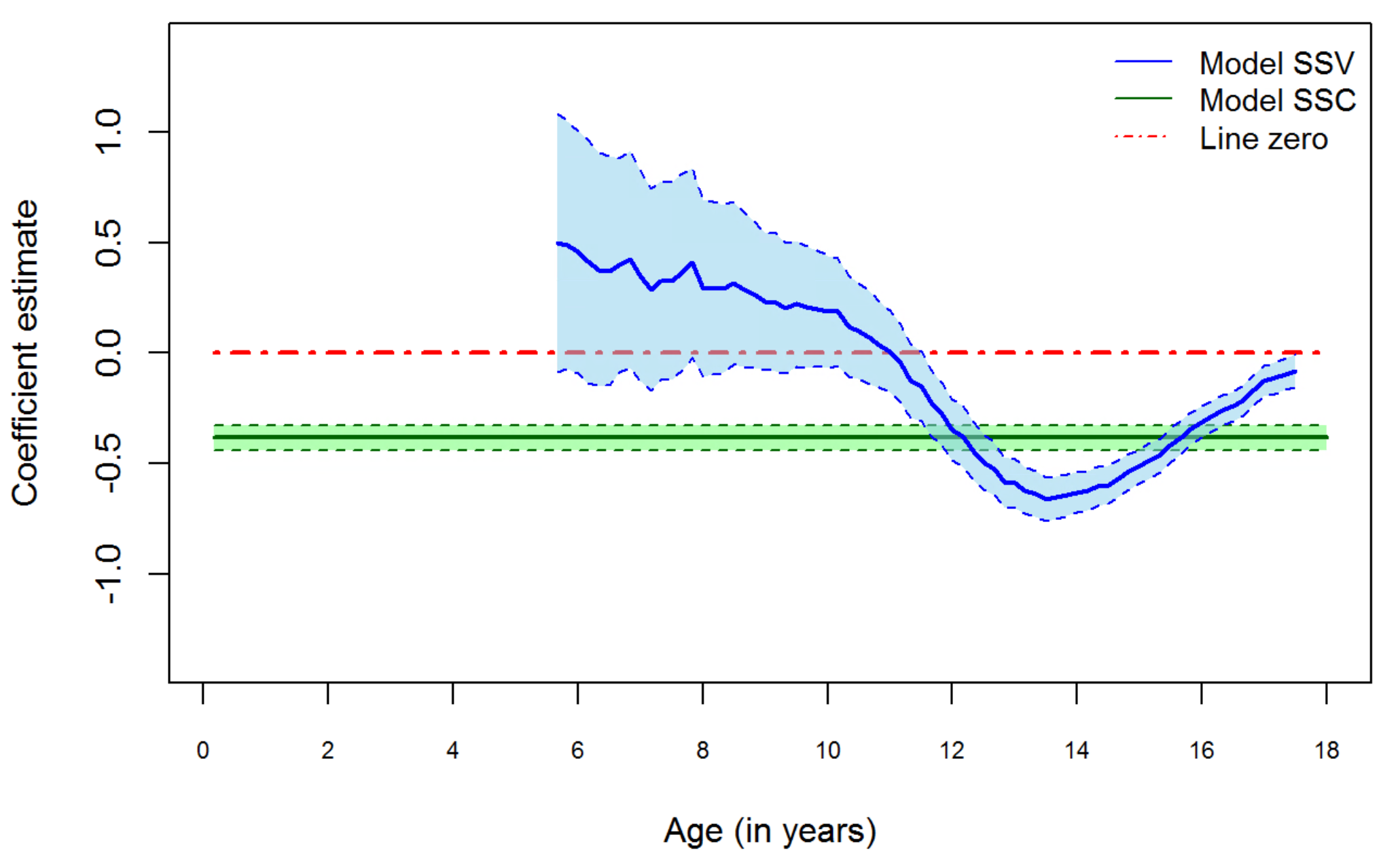}
\caption{Sex (male vs. female); stratum 2} \label{fig:R_b}
\end{subfigure}

\caption{Estimated regression coefficients for males with 95\% pointwise confidence intervals based on MHED data.\\ \footnotesize{\noindent Note: Model (SSV): $\lambda(a\mid \mathcal{H}_i(a), Z_i)=\lambda_{0s}(a) \exp\{\beta_s(a)^{'} Z_i\}$; Model (SSC): $\lambda(a\mid \mathcal{H}_i(a), $ $Z_i)=\lambda_{0s}(a) \exp\{\beta_s' Z_i\}$.}} \label{fig:R_coef_sex}
\end{figure}

\begin{figure}[t!] 
\begin{subfigure}{0.48\textwidth}
\includegraphics[width=\linewidth]{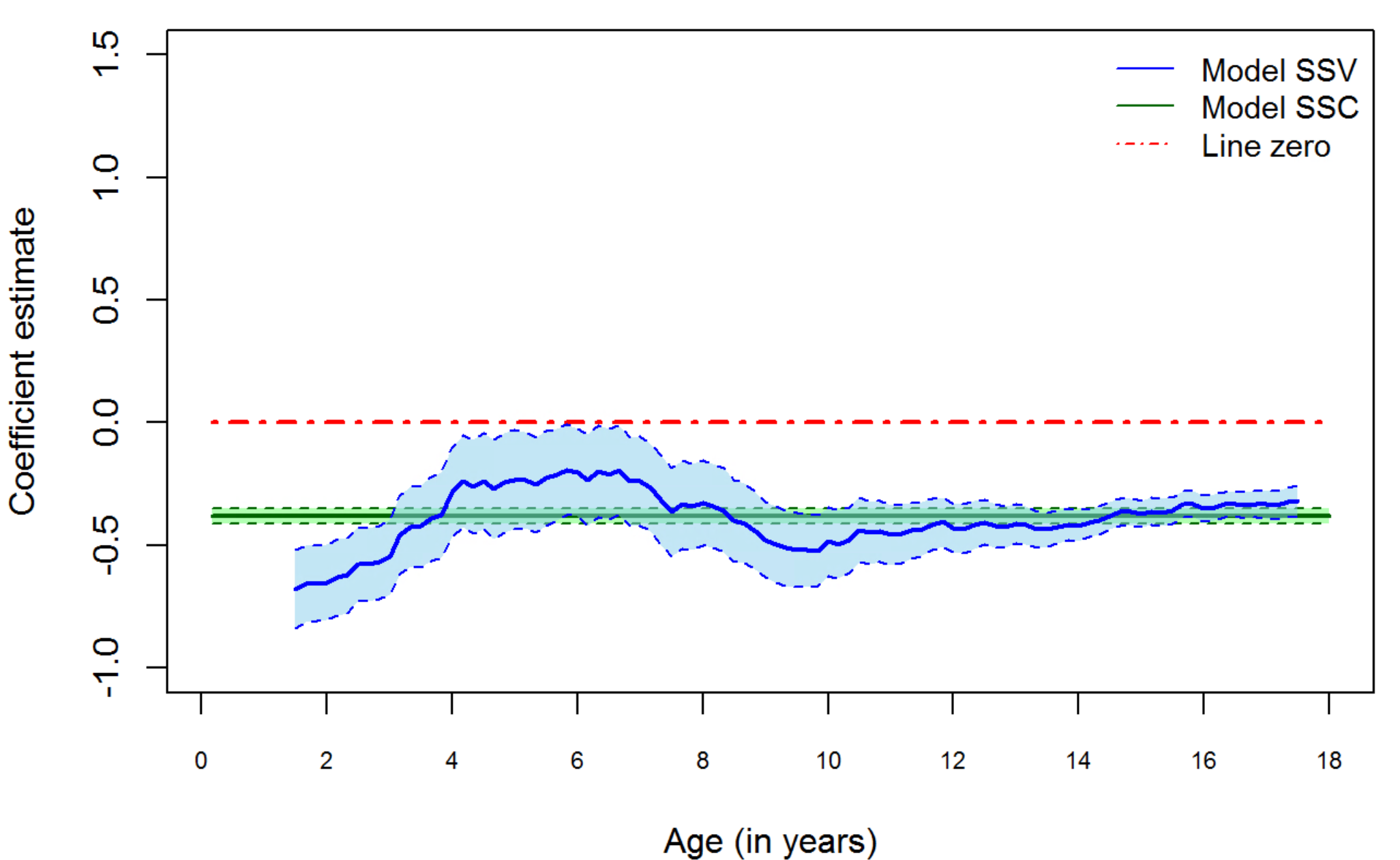}
\caption{Edmonton (vs. other regions); stratum 1} \label{fig:R_c}
\end{subfigure}\hspace*{\fill}
\begin{subfigure}{0.48\textwidth}
\includegraphics[width=\linewidth]{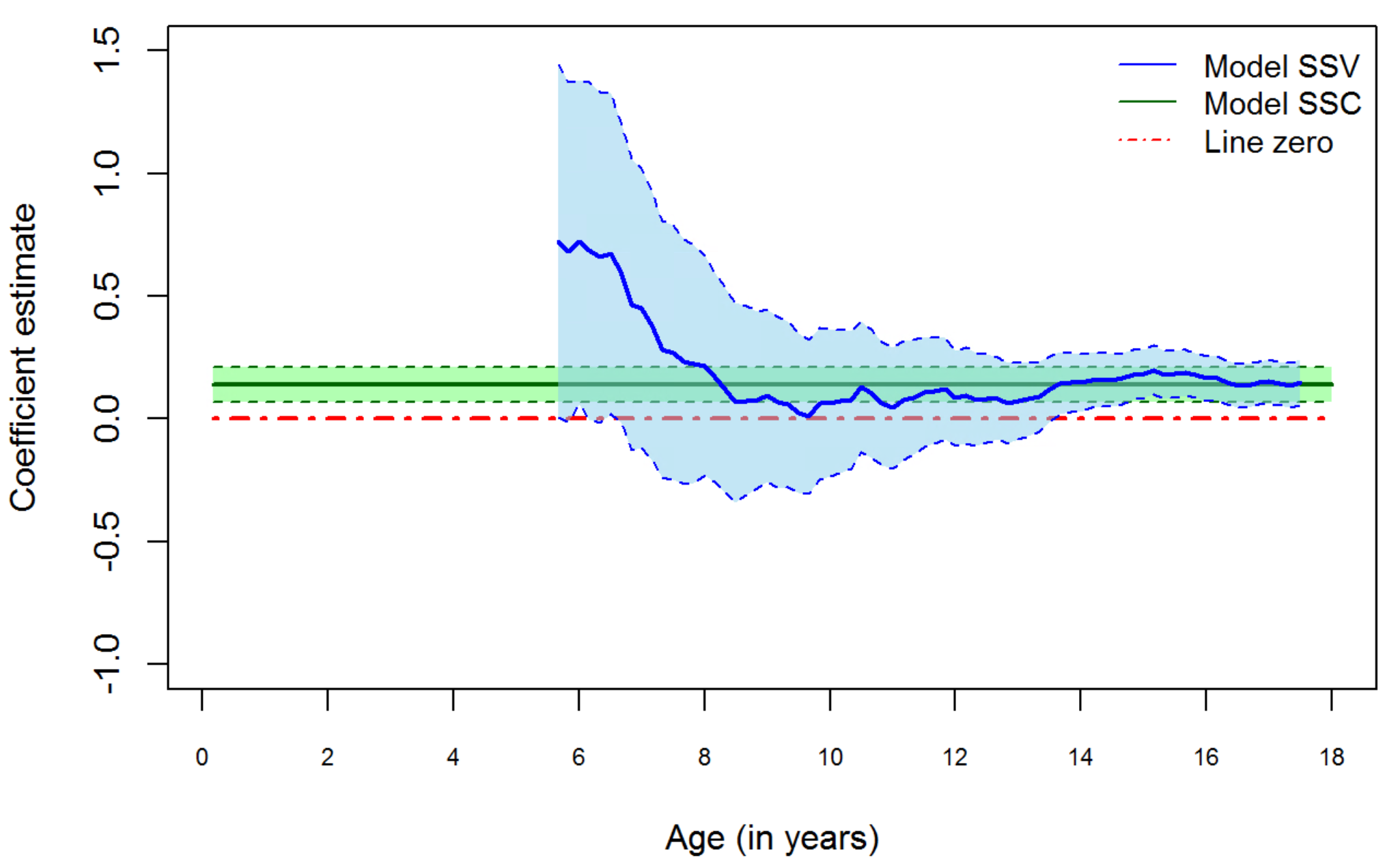}
\caption{Edmonton (vs. other regions); stratum 2} \label{fig:R_d}
\end{subfigure}

\medskip
\begin{subfigure}{0.48\textwidth}
\includegraphics[width=\linewidth]{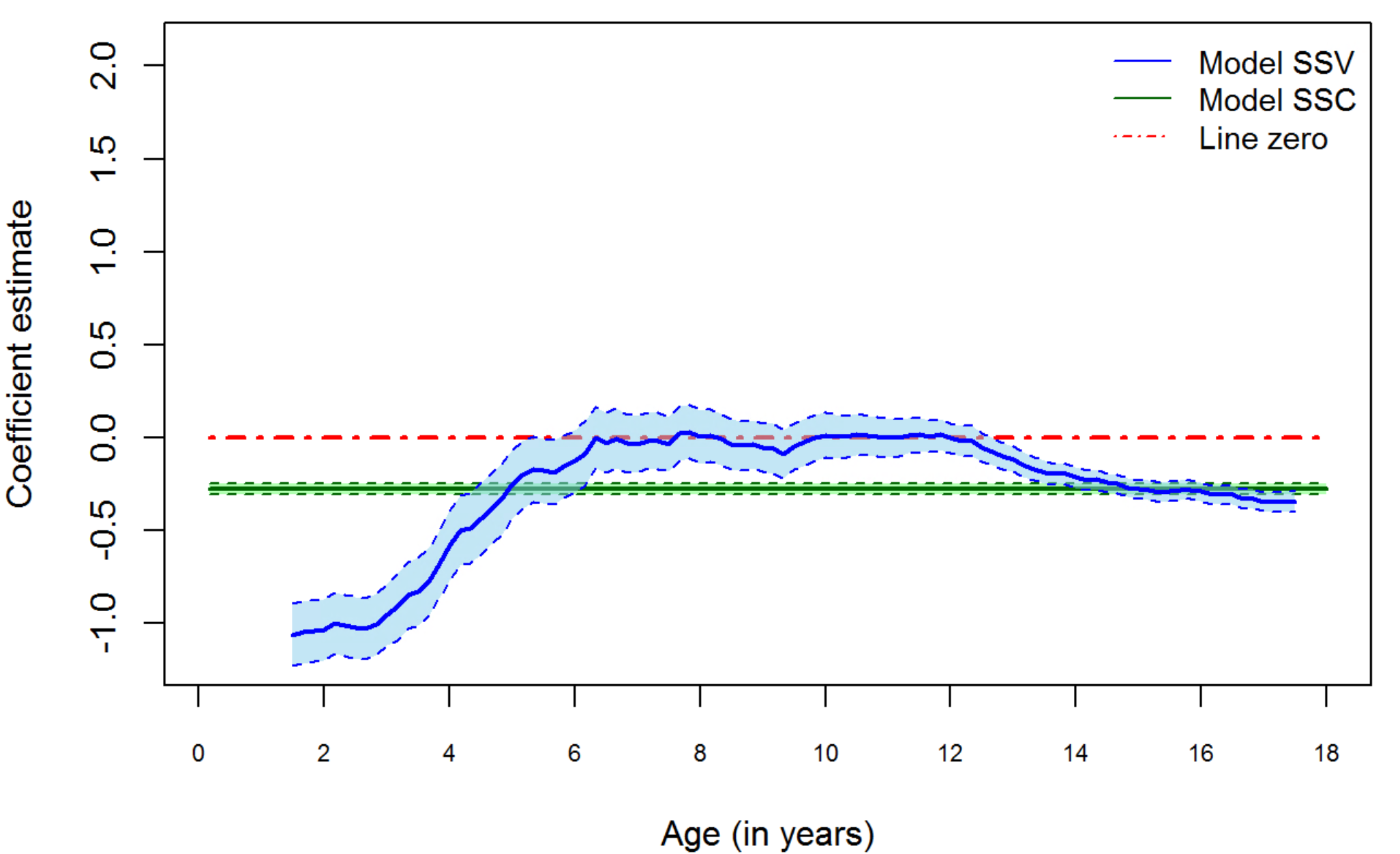}
\caption{Calgary (vs. other regions); stratum 1} \label{fig:R_e}
\end{subfigure}\hspace*{\fill}
\begin{subfigure}{0.48\textwidth}
\includegraphics[width=\linewidth]{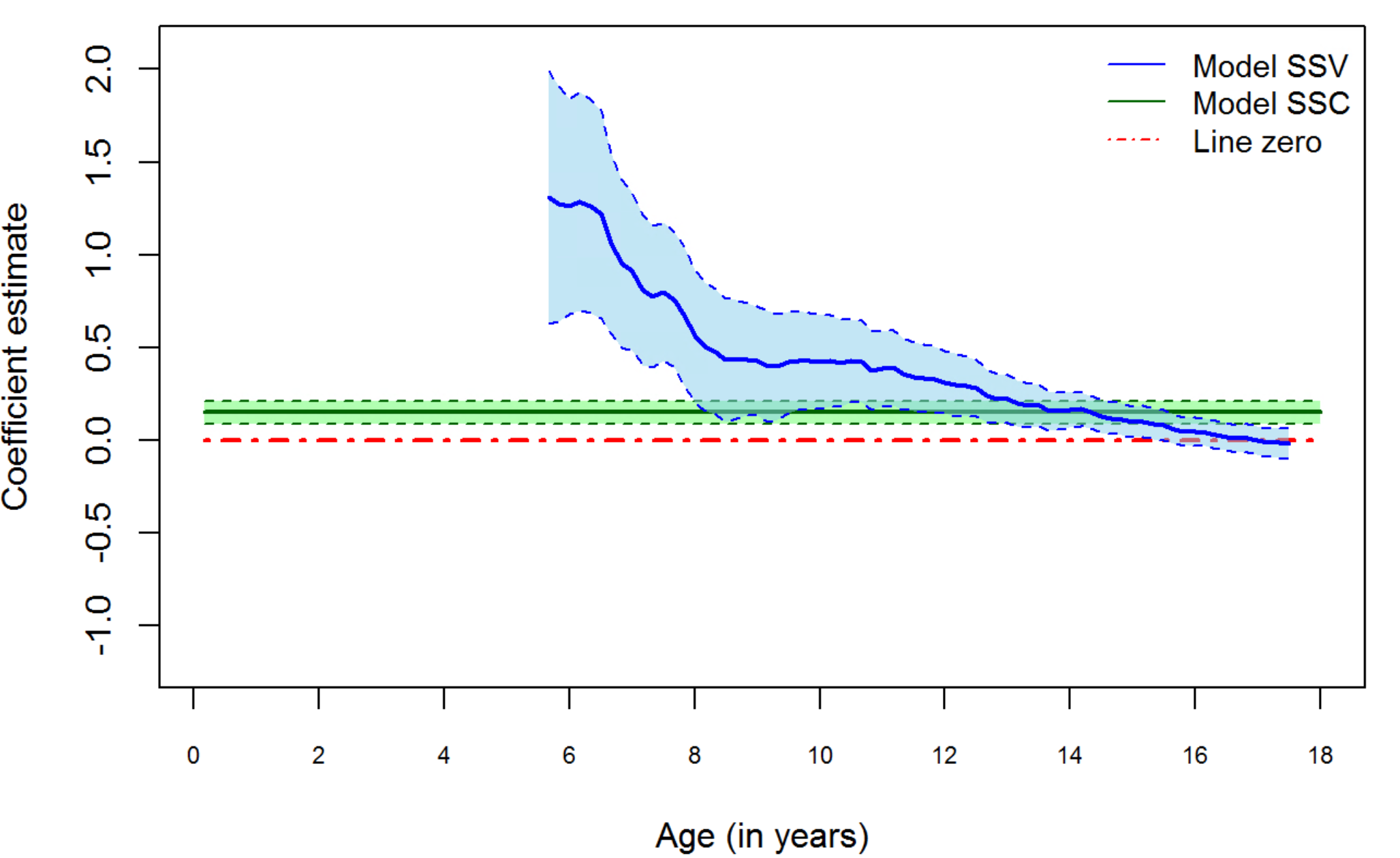}
\caption{Calgary (vs. other regions); stratum 2} \label{fig:R_f}
\end{subfigure}

\caption{Estimated regression coefficients for regions with 95\% pointwise confidence intervals based on MHED data.\\ \footnotesize{\noindent Note: Model (SSV): $\lambda(a\mid \mathcal{H}_i(a), Z_i)=\lambda_{0s}(a) \exp\{\beta_s(a)^{'} Z_i\}$; Model (SSC): $\lambda(a\mid \mathcal{H}_i(a), $ $Z_i)=\lambda_{0s}(a) \exp\{\beta_s' Z_i\}$.}} \label{fig:R_coef_region}
\end{figure}

\begin{figure}[!ht]
\centering
\includegraphics[width=0.8\textwidth]{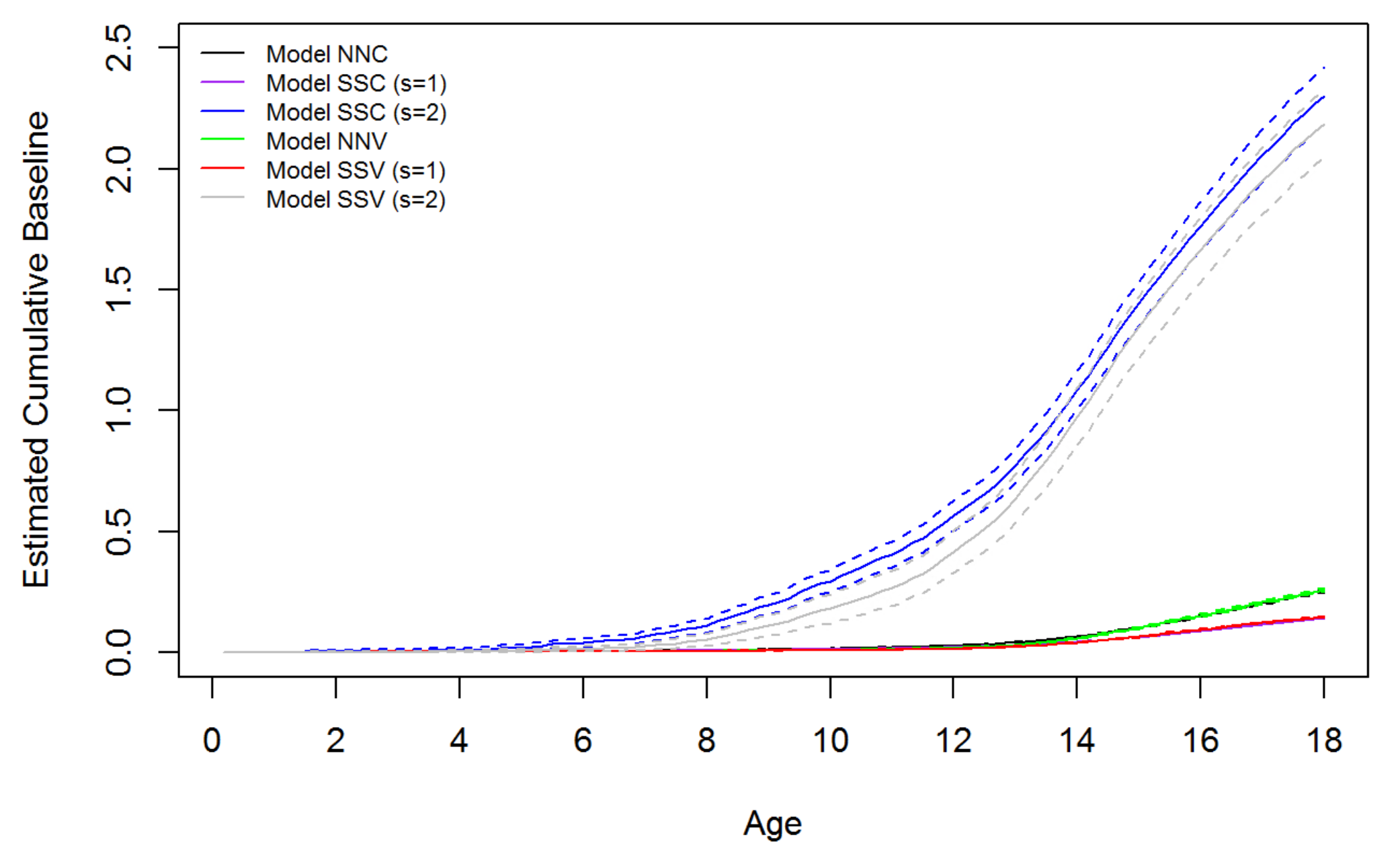}
\caption{\label{Figure:real_exa_cum_baseline}Estimated cumulative baseline intensity functions with 95\% pointwise confidence intervals based on MHED data.\\ \footnotesize{\noindent Note: Model (SSV): $\lambda(a\mid \mathcal{H}_i(a), Z_i)=\lambda_{0s}(a) \exp\{\beta_s(a)^{'} Z_i\}$; Model (SSC): $\lambda(a\mid \mathcal{H}_i(a), $ $Z_i)=\lambda_{0s}(a) \exp\{\beta_s' Z_i\}$; Model (NNV): $\lambda(a\mid \mathcal{H}_i(a), Z_i)=\lambda_{0}(a) \exp\{\beta(a)^{'} Z_i\}$;  Model (NNC): $\lambda(a\mid \mathcal{H}_i(a), Z_i)=\lambda_{0}(a) \exp\{\beta' Z_i\}$.}}
\end{figure}

We presented the main estimation results under Models (\ref{eq:model} or SSV) and (SSC). Figures \ref{fig:R_coef_sex} and \ref{fig:R_coef_region} show the estimated regression coefficients under Models (SSV) and (SSC) with 95\% pointwise confidence intervals. Both resampling-based and bootstrap-based variance estimates were computed with $1,000$ repetitions and gave us very similar results. Our exploration suggested that 400 realizations of multipliers were sufficient for the resampling method in the MHED example. The variance estimates shown in the figure are based on $1,000$ bootstrap samples. In Figures \ref{fig:R_coef_sex} and \ref{fig:R_coef_region}, the blue solid curves denote the estimated time-varying coefficients under Model (SSV) and the green solid lines are the estimated time-independent coefficients under Model (SSC). The red dashed lines indicate zero. Figure (\ref{fig:R_a}) suggests that boys were more likely than girls from the same region to have their first MHED visit before age $11$, but were less likely to experience their first visit during adolescence. Figure (\ref{fig:R_b}) indicates that teenage boys had a lower risk of making additional MHED visits than girls after their first visit. The estimated time-varying coefficient was truncated before age $5.67$ (5 years and 8 months old) due to a lack of data at younger ages. Figure \ref{fig:R_coef_region} shows that most segments of the estimated coefficients are below zero in stratum $1$ and above zero in stratum $2$. These patterns suggest that residents of Edmonton and Calgary were less likely to have the first MHED visit but more likely to have subsequent visits compared to those of other regions with the same sex. One possible explanation is that individuals in large cities might initially have a lower risk of MHED visits; however, once they sought care for a mental health crisis and received treatment in EDs, they might be more likely to return to the EDs for similar issues. Notably, Calgary and Edmonton are the only two locations in Alberta with pediatric EDs. Furthermore, the estimated coefficients under Model (SSV) vary over time, emphasizing the importance of accounting for time-varying coefficients. Variations in covariate effects across strata indicate the need for stratification in the regression coefficients.

Figure \ref{Figure:real_exa_cum_baseline} presents the estimated cumulative baselines under different models. The significant differences in the estimated cumulative baselines across strata highlight the need for stratification in the baseline intensity functions. Overall, the results of the real data example support our proposed model.

\section{Simulation Study}\label{sec:simulation}
We conducted a simulation study to evaluate the finite-sample performance of our proposed approach in Section \ref{sec:estimation}. We simulated a target population $\mathcal{P}$ of 200,000 subjects and selected those with at least one event within a 7- or 18-year window to form the study cohort (around 10\% of the population). We considered three indicator variables, denoted as $Z_1$, $Z_2$, and $Z_3$, as covariates. The indicator $Z_1$ followed a Bernoulli distribution with a probability of 0.5. A continuous variable $X$ was generated from a lognormal distribution with its natural logarithm having a mean of $log(8)$ and a variance of $log(3)^2$. The indicators $Z_2$ and $Z_3$ were defined as $Z_{2}=I(5<X\leq 13)$ and $Z_{3}=I(X>13)$. The stratification variable (\ref{eq:stratification_variable}) and the Epanechnikov kernel were used in the simulation study. We set $\tau_L=6$ units (1 year old), and other settings were the same with those used in the real data example. We conducted $1,000$ simulation repetitions.

We considered two simulation scenarios, each using a different true model to generate event data. The true models for each scenario are listed below.
\begin{itemize}
\item \textbf{Scenario 1}: $\lambda(a\mid \mathcal{H}_i(a), Z_i)
=\lambda_{0}(a)\exp\{\beta(a)^{'}Z_{i}\}$ with $Z_i=(Z_{i1},Z_{i2},$ $ Z_{i3})^{'}$ for $i \in \mathcal{P}$; that is Model (NNV).
\item \textbf{Scenario 2}: $\lambda(a\mid \mathcal{H}_i(a), Z_i)
=\lambda_{0s}(a)\exp\{\beta_s(a)^{'}Z_{i}\}$ for $i \in \mathcal{P}$ and $s\in\{1, 2\}$; that is Model (\ref{eq:model} or SSV).
\end{itemize}
In each scenario, we conducted analyses on the generated event data under the models in Table \ref{Tab:submodels}. In Scenario 1, the true model is a relatively simple model without stratification, while our proposed model is more complex. This allows us to assess the consistency and efficiency of our proposed estimators. In Scenario 2, the true model is our proposed model, so we can verify the consistency of the estimators. For the simpler models, we could evaluate the robustness of the estimators when the models are mis-specified. In the remainder of this section, we focused on the simulation results for Scenario 2 with a 7-year data extraction window.

\medskip

\noindent5.1\textit{\hspace{0.4cm}Results of Simulation Scenario 2} \label{sec:simulation_sce2}

\begin{figure}[t!] 
\begin{subfigure}{0.48\textwidth}
\includegraphics[width=\linewidth]{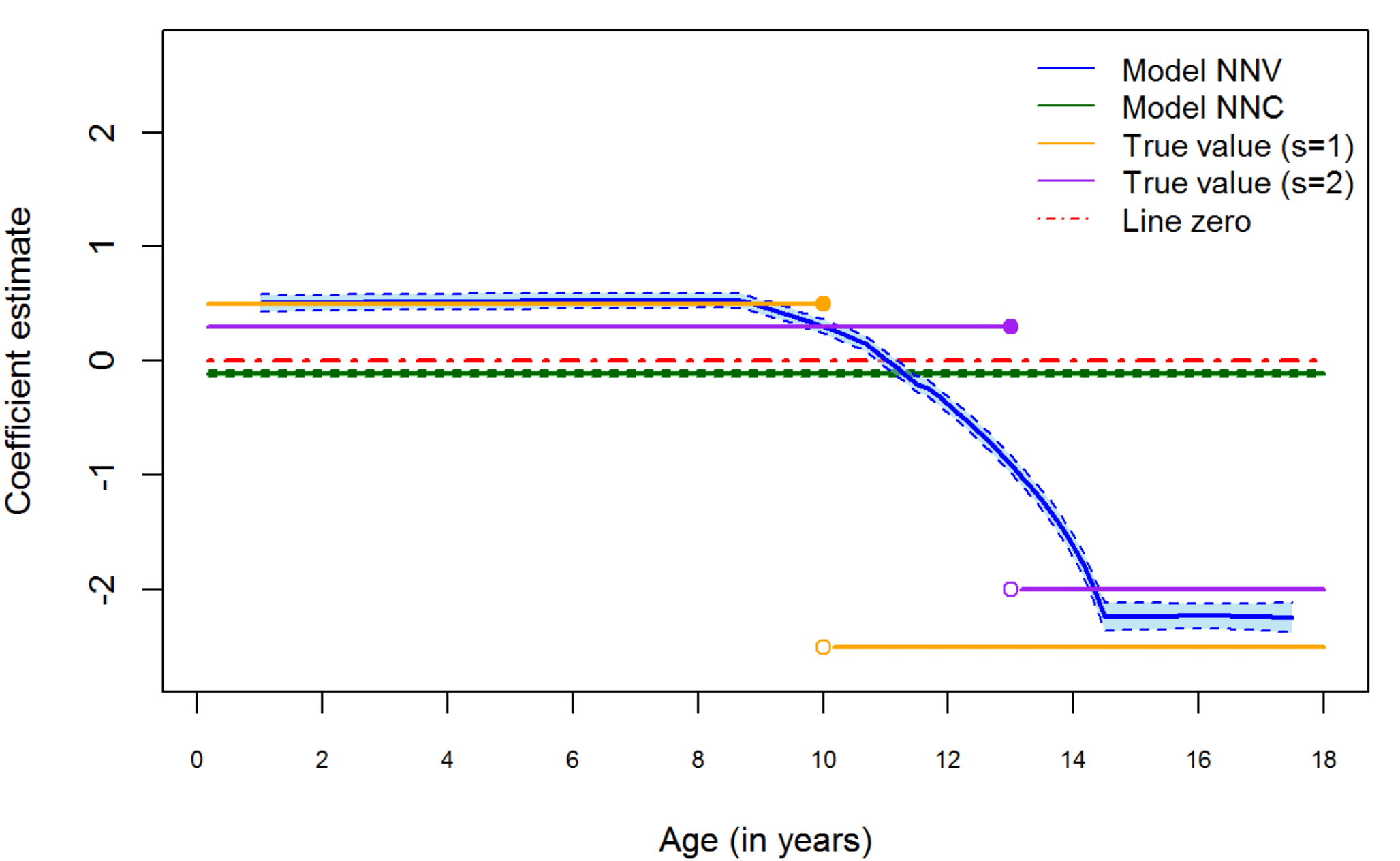}
\caption{Under Model (NNV)} \label{fig:S_a}
\end{subfigure}\hspace*{\fill}
\begin{subfigure}{0.48\textwidth}
\includegraphics[width=\linewidth]{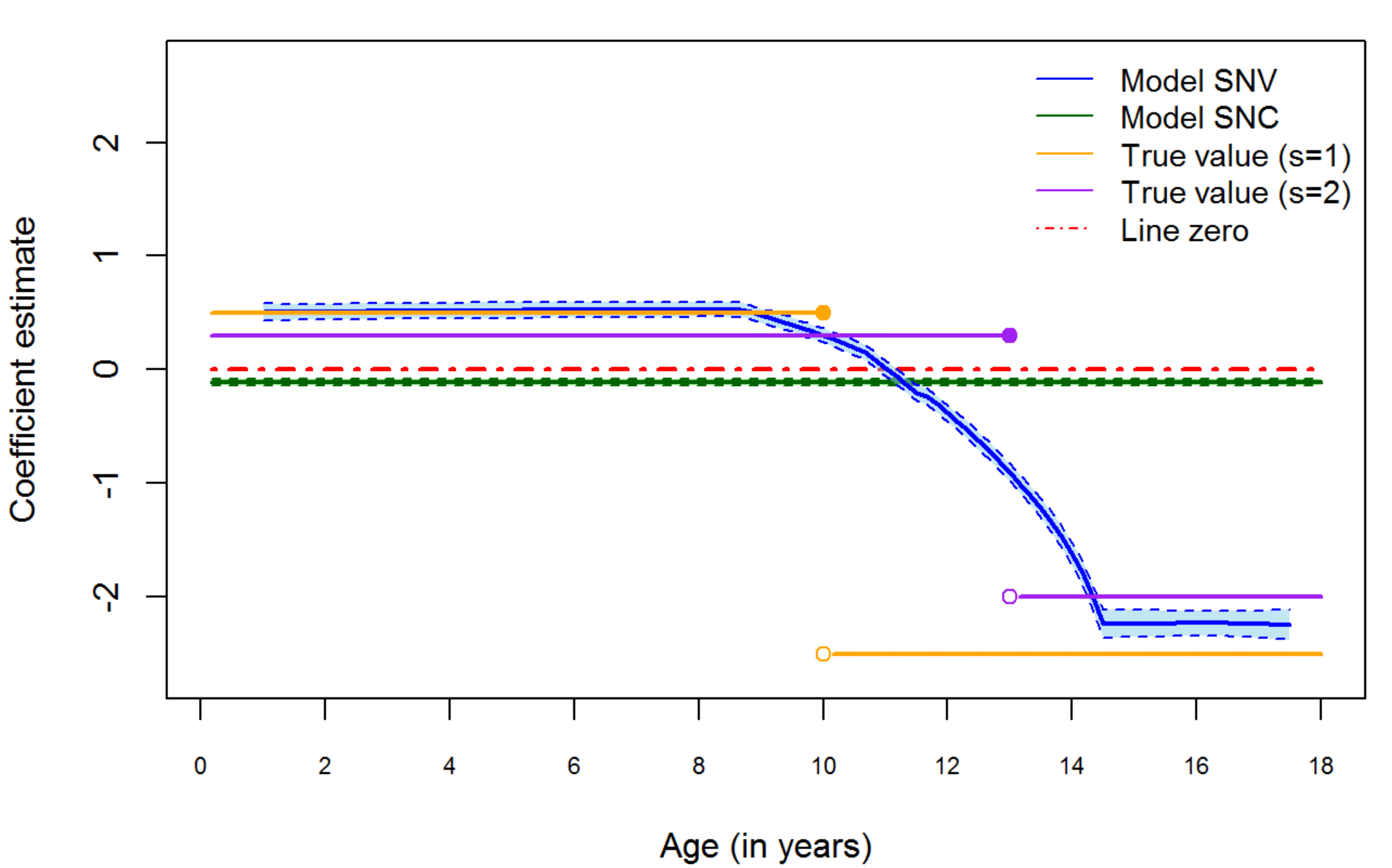}
\caption{Under Model (SNV)} \label{fig:S_b}
\end{subfigure}

\medskip
\begin{subfigure}{0.48\textwidth}
\includegraphics[width=\linewidth]{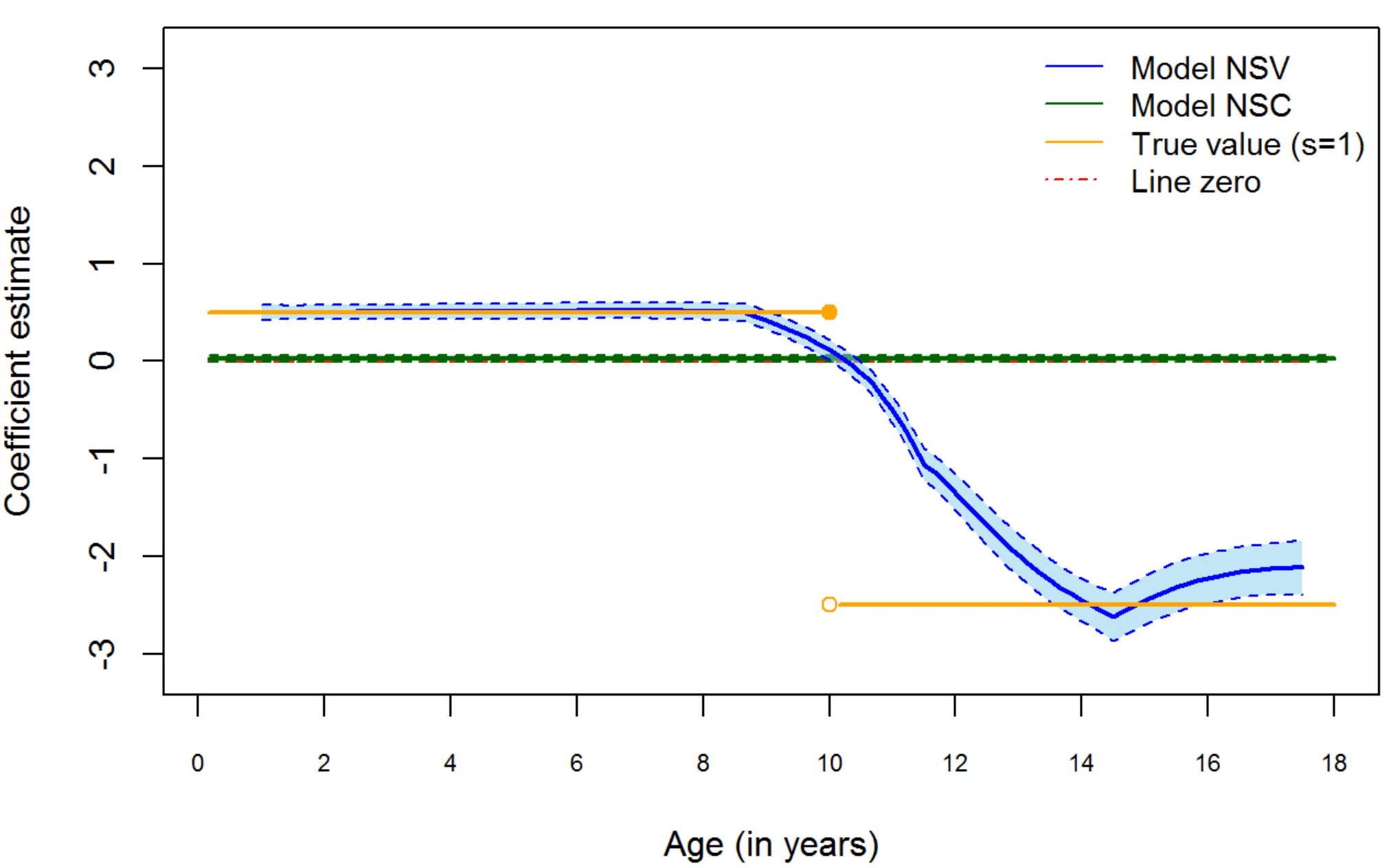}
\caption{Under Model (NSV); stratum 1} \label{fig:S_c}
\end{subfigure}\hspace*{\fill}
\begin{subfigure}{0.48\textwidth}
\includegraphics[width=\linewidth]{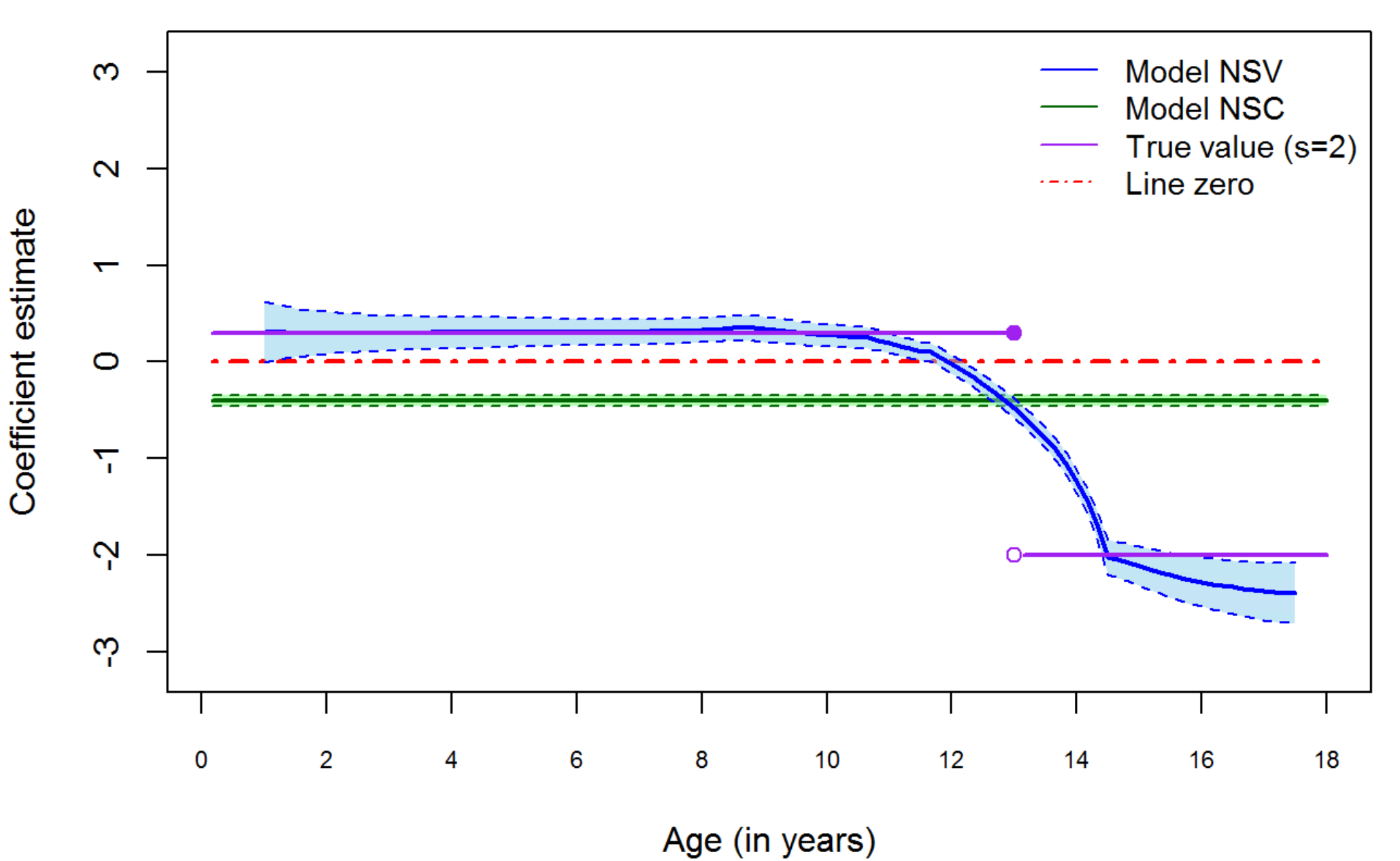}
\caption{Under Model (NSV); stratum 2} \label{fig:S_d}
\end{subfigure}

\medskip
\begin{subfigure}{0.48\textwidth}
\includegraphics[width=\linewidth]{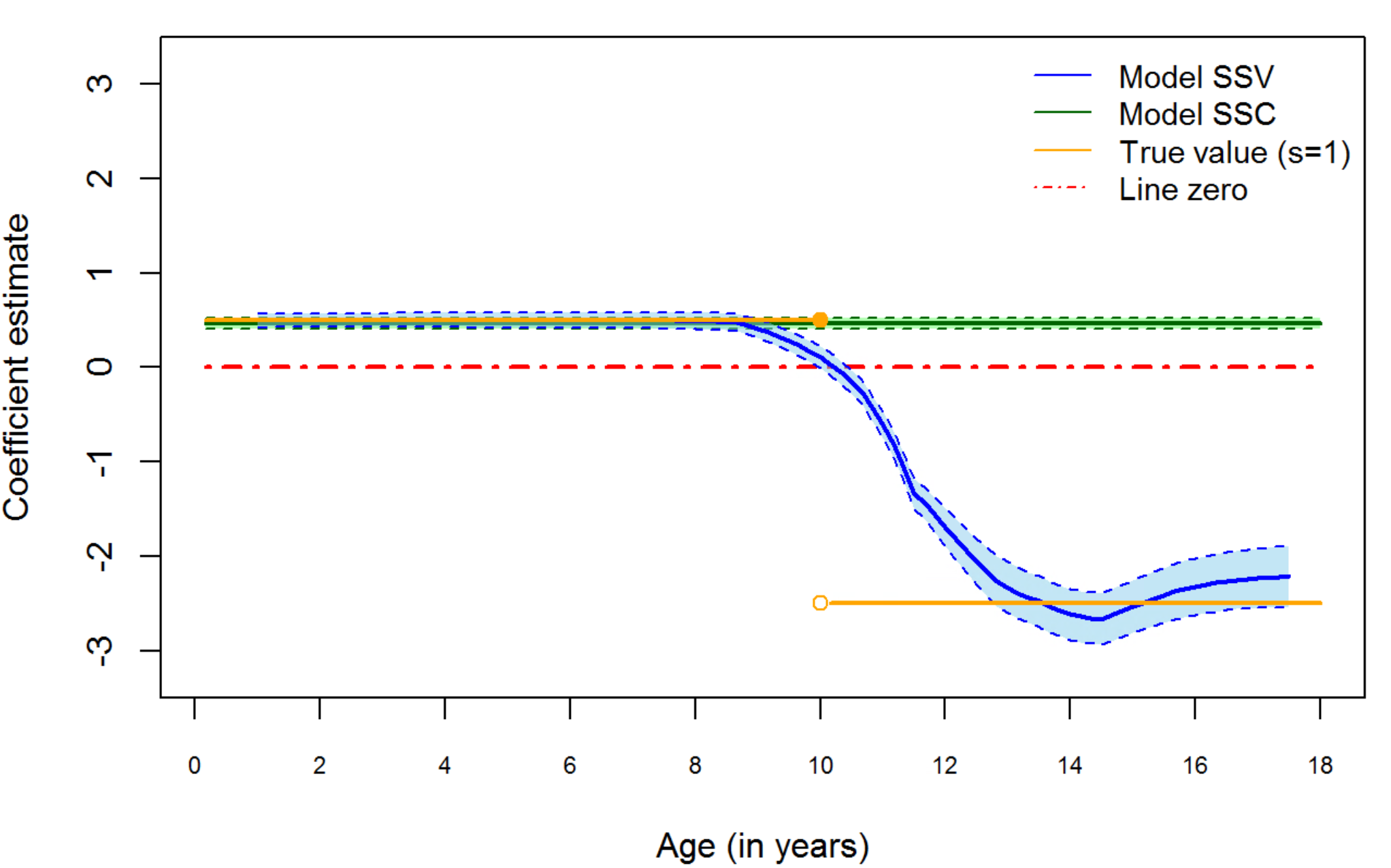}
\caption{Under Model (SSV); stratum 1} \label{fig:S_e}
\end{subfigure}\hspace*{\fill}
\begin{subfigure}{0.48\textwidth}
\includegraphics[width=\linewidth]{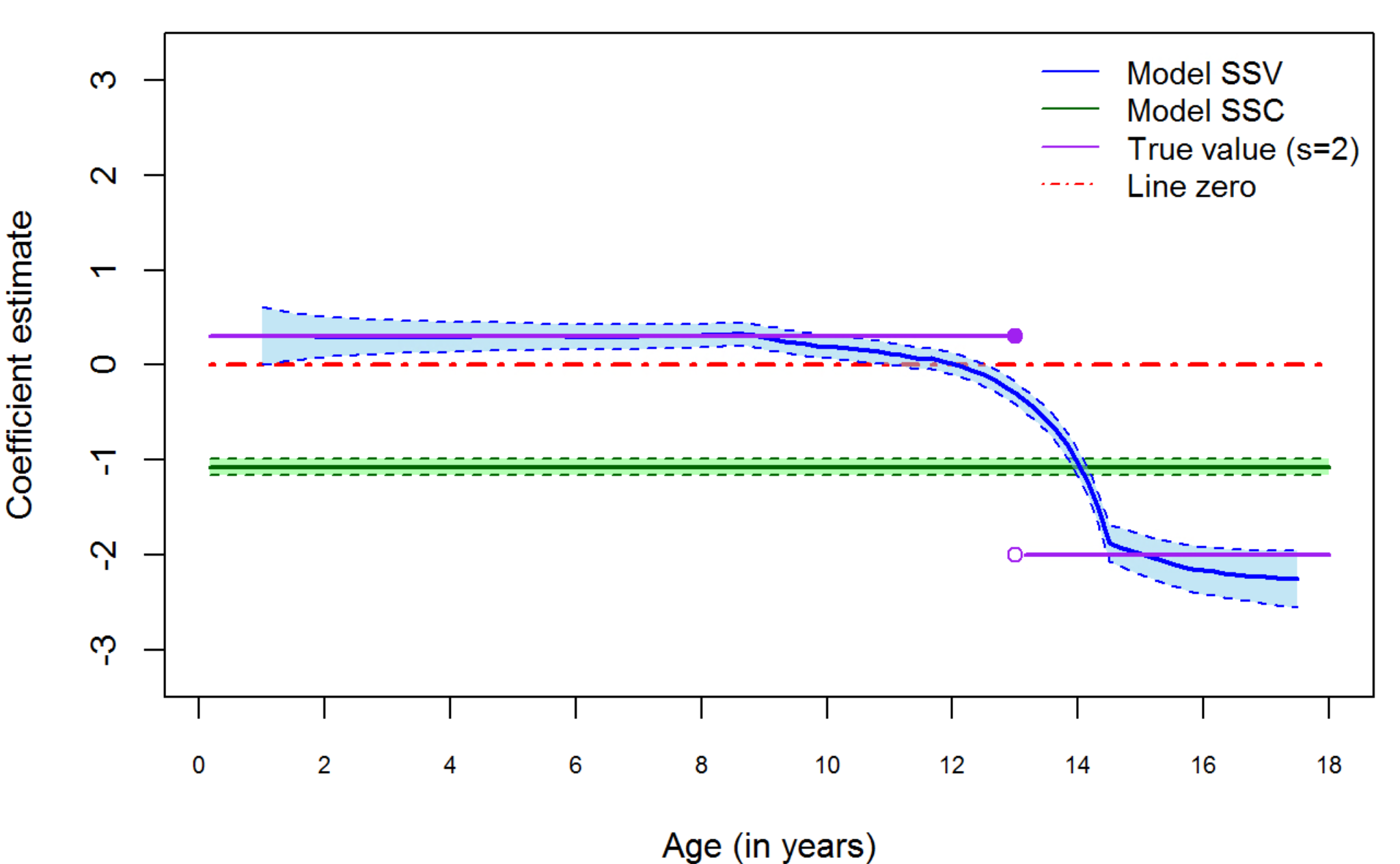}
\caption{Under Model (SSV); stratum 2} \label{fig:S_f}
\end{subfigure}


\caption{Estimated coefficients to the indicator $Z_1$ with 95\% pointwise confidence intervals in Scenario 2 using a 7-year data extraction window based on $1,000$ simulation repetitions.\\ \footnotesize{\noindent Note: The models are shown in Table (\ref{Tab:submodels}).}} \label{fig:S_coef}
\end{figure}

\begin{figure}[!htb] 
\begin{subfigure}{0.48\textwidth}
\includegraphics[width=\linewidth]{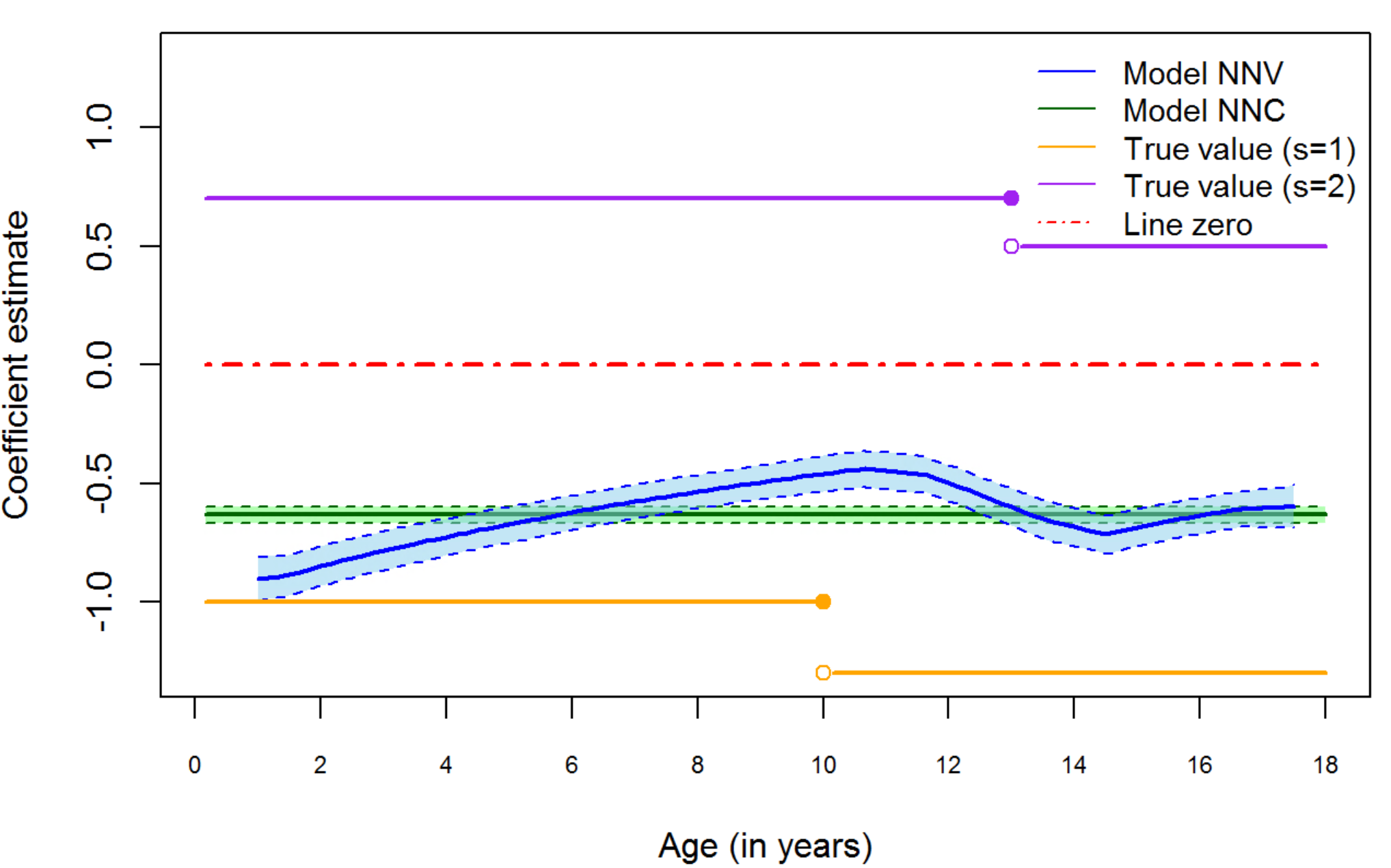}
\caption{Under Model (NNV)} 
\end{subfigure}\hspace*{\fill}
\begin{subfigure}{0.48\textwidth}
\includegraphics[width=\linewidth]{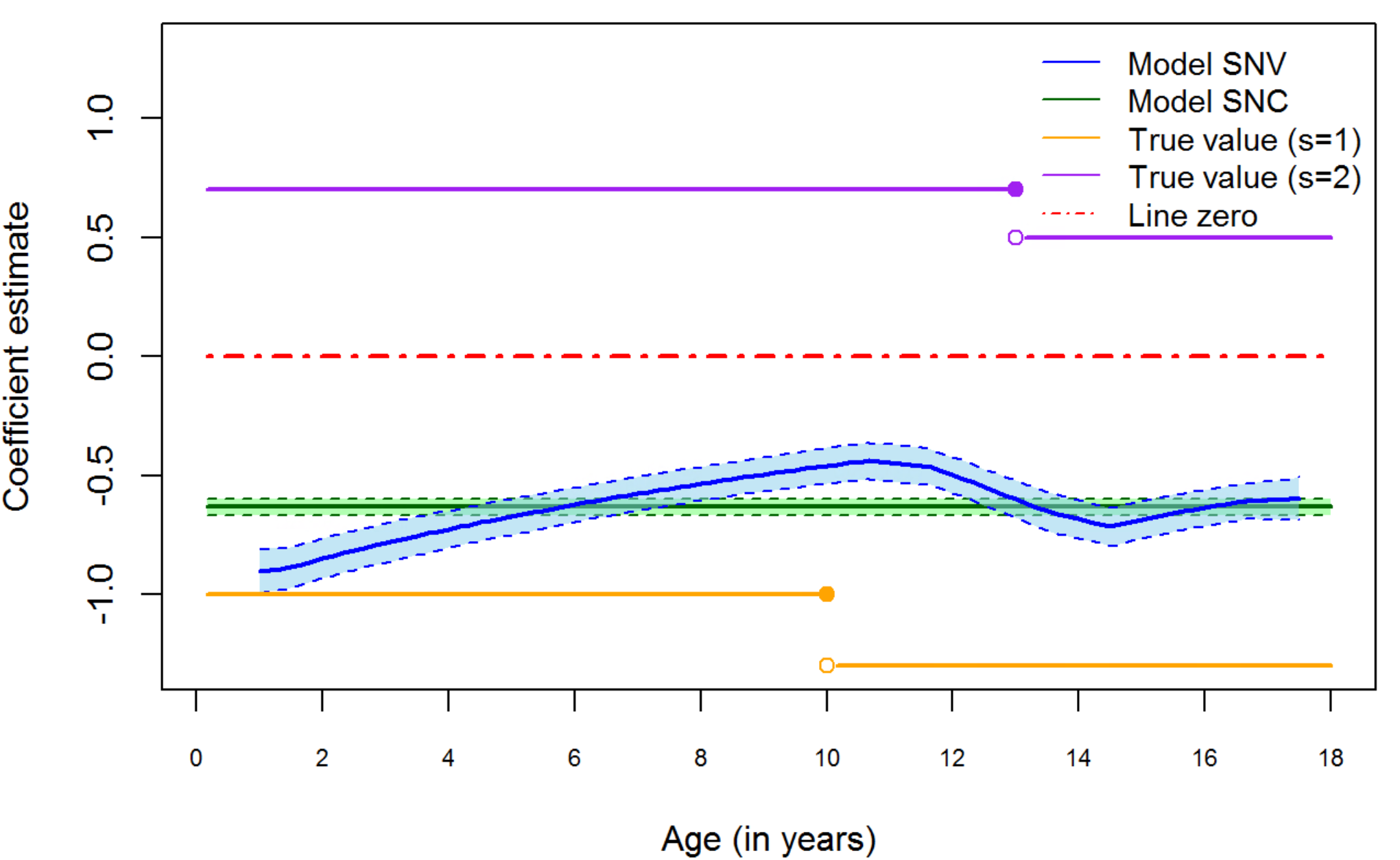}
\caption{Under Model (SNV)} 
\end{subfigure}

\medskip
\begin{subfigure}{0.48\textwidth}
\includegraphics[width=\linewidth]{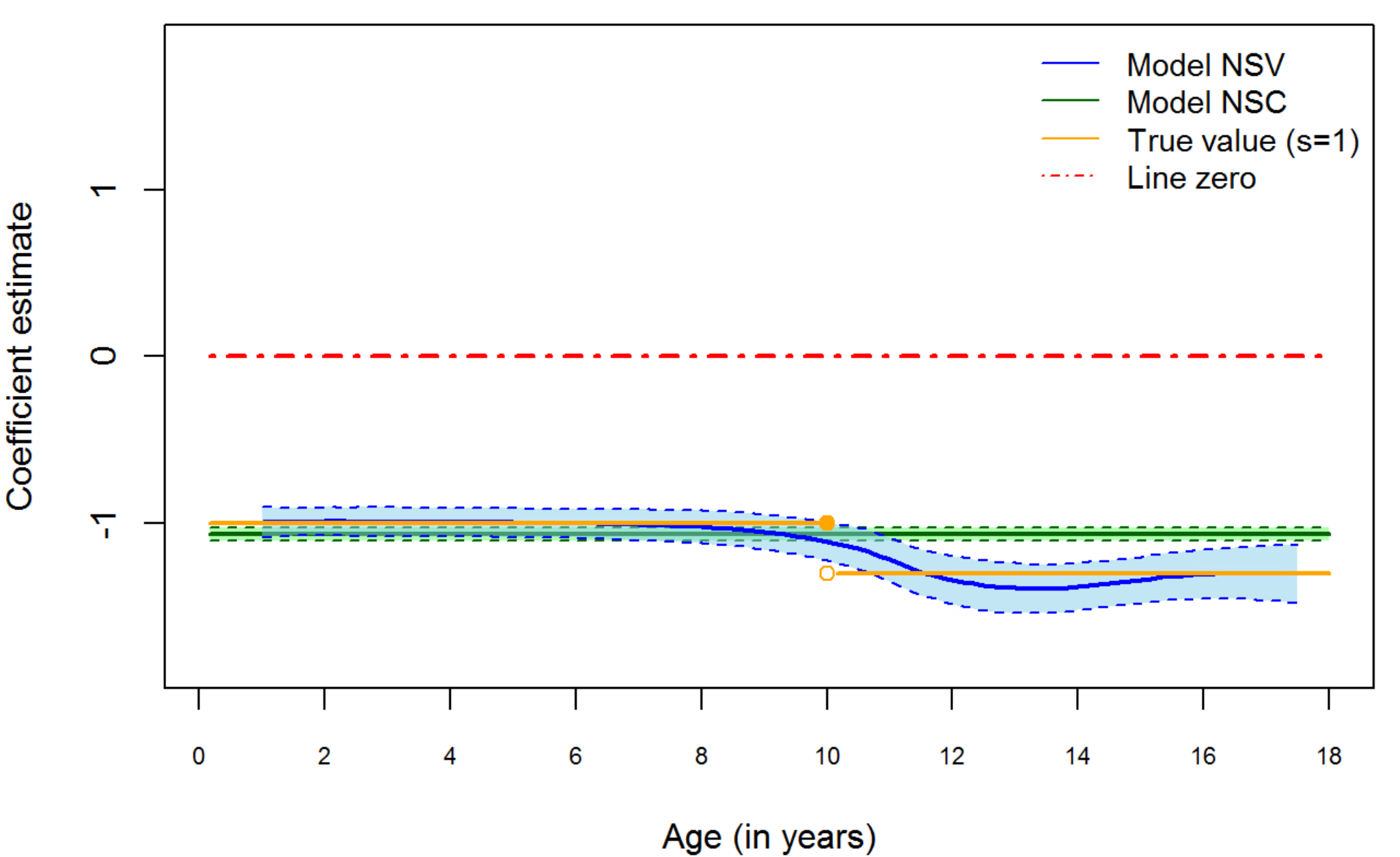}
\caption{Under Model (NSV); stratum 1} 
\end{subfigure}\hspace*{\fill}
\begin{subfigure}{0.48\textwidth}
\includegraphics[width=\linewidth]{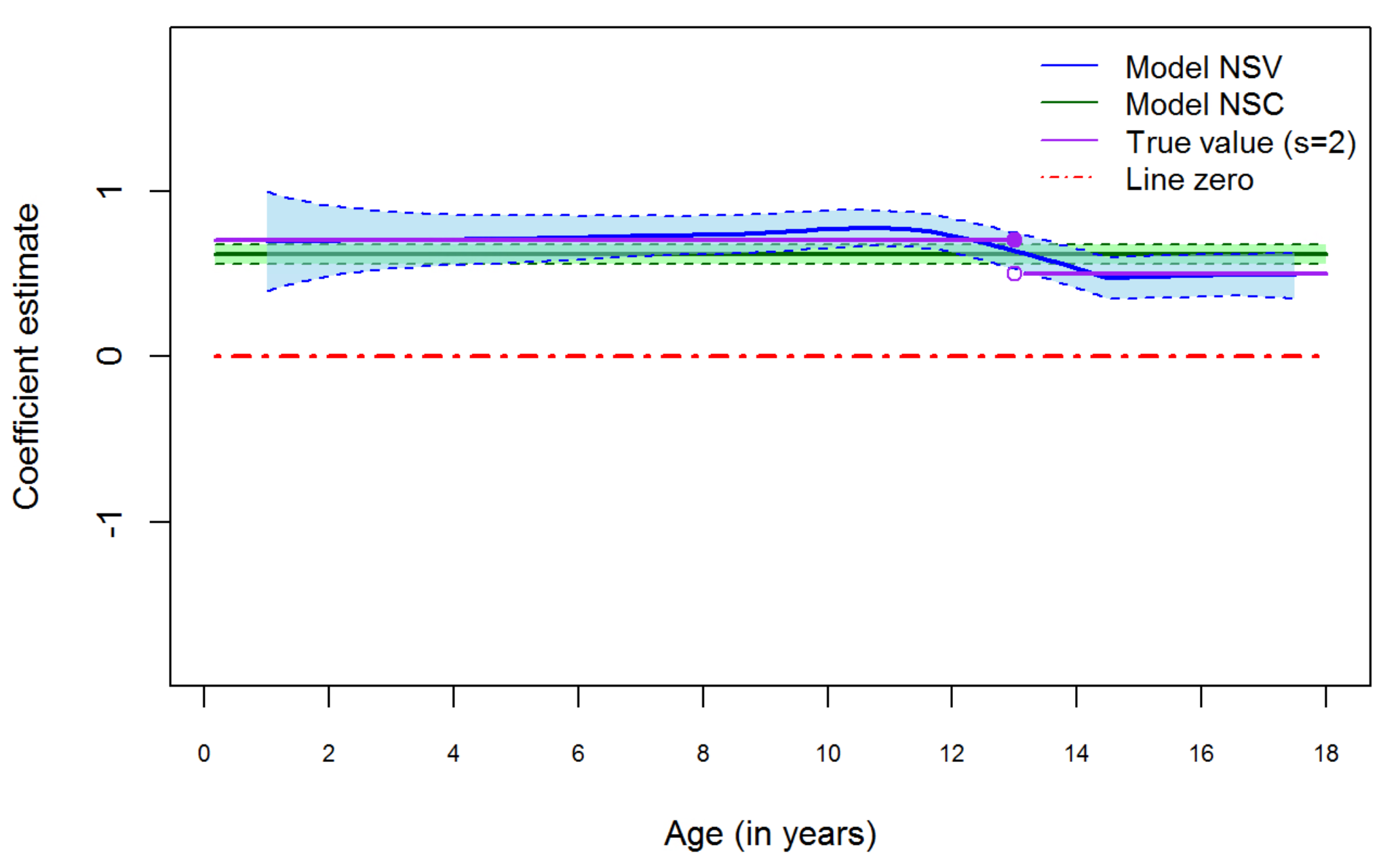}
\caption{Under Model (NSV); stratum 2} 
\end{subfigure}

\medskip
\begin{subfigure}{0.48\textwidth}
\includegraphics[width=\linewidth]{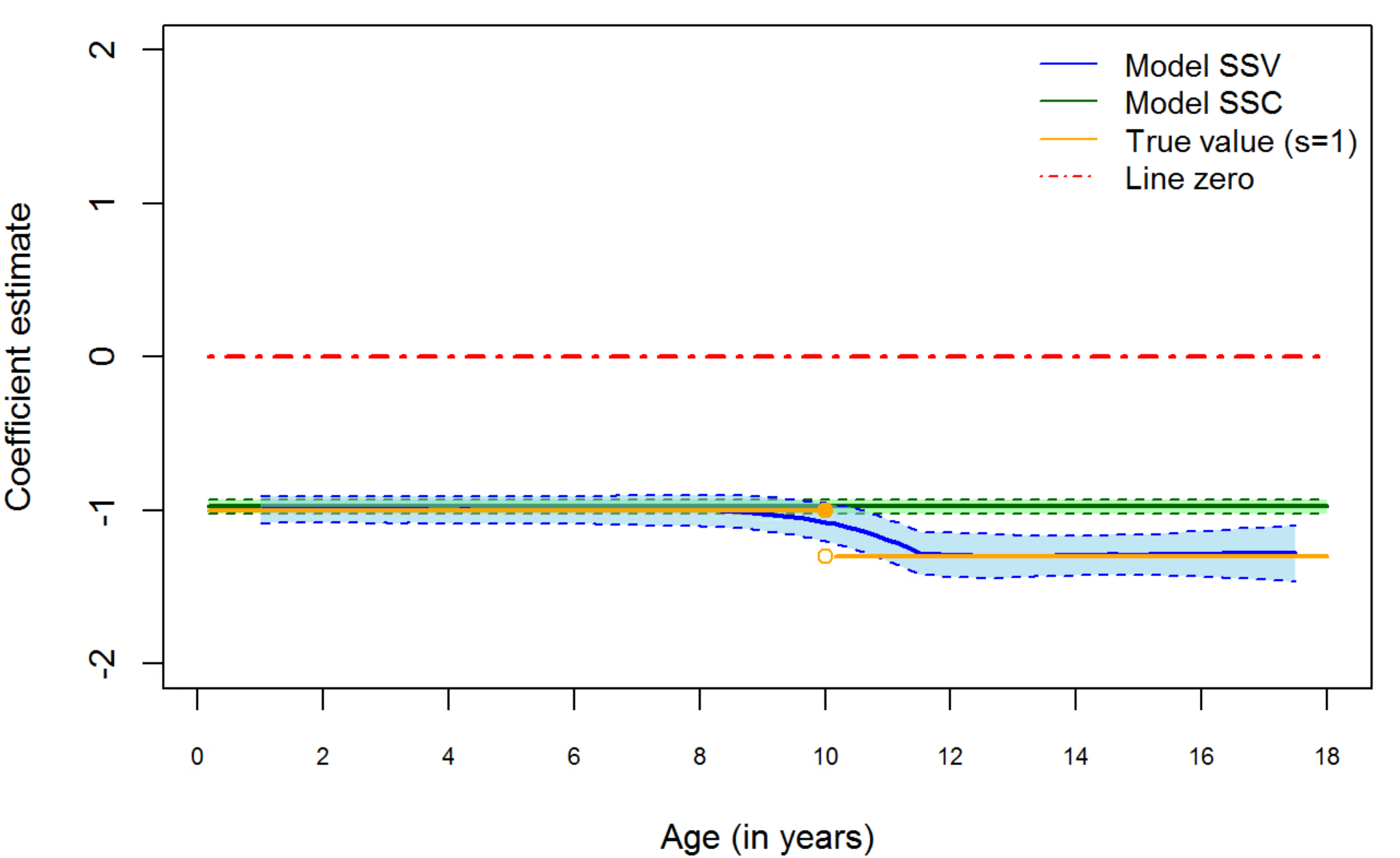}
\caption{Under Model (SSV); stratum 1}
\end{subfigure}\hspace*{\fill}
\begin{subfigure}{0.48\textwidth}
\includegraphics[width=\linewidth]{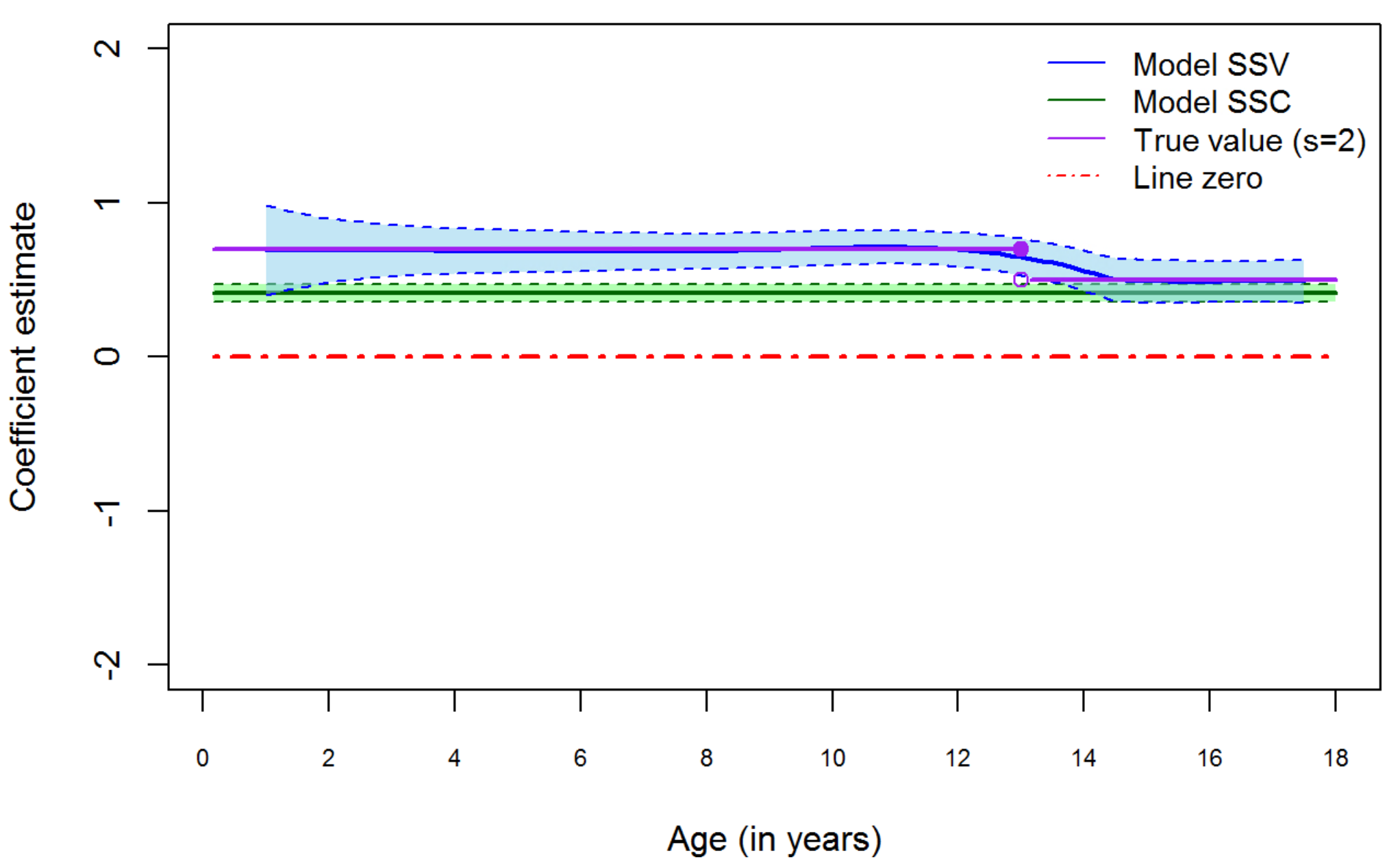}
\caption{Under Model (SSV); stratum 2} 
\end{subfigure}

\caption{Estimated coefficients to the indicator $Z_2$ with 95\% pointwise confidence intervals in Scenario 2 using a 7-year data extraction window based on $1,000$ simulation repetitions.\\ \footnotesize{\noindent Note: The models are shown in Table (\ref{Tab:submodels}).}} \label{fig:S_coef_z2}
\end{figure}

\begin{figure}[!htb] 
\begin{subfigure}{0.48\textwidth}
\includegraphics[width=\linewidth]{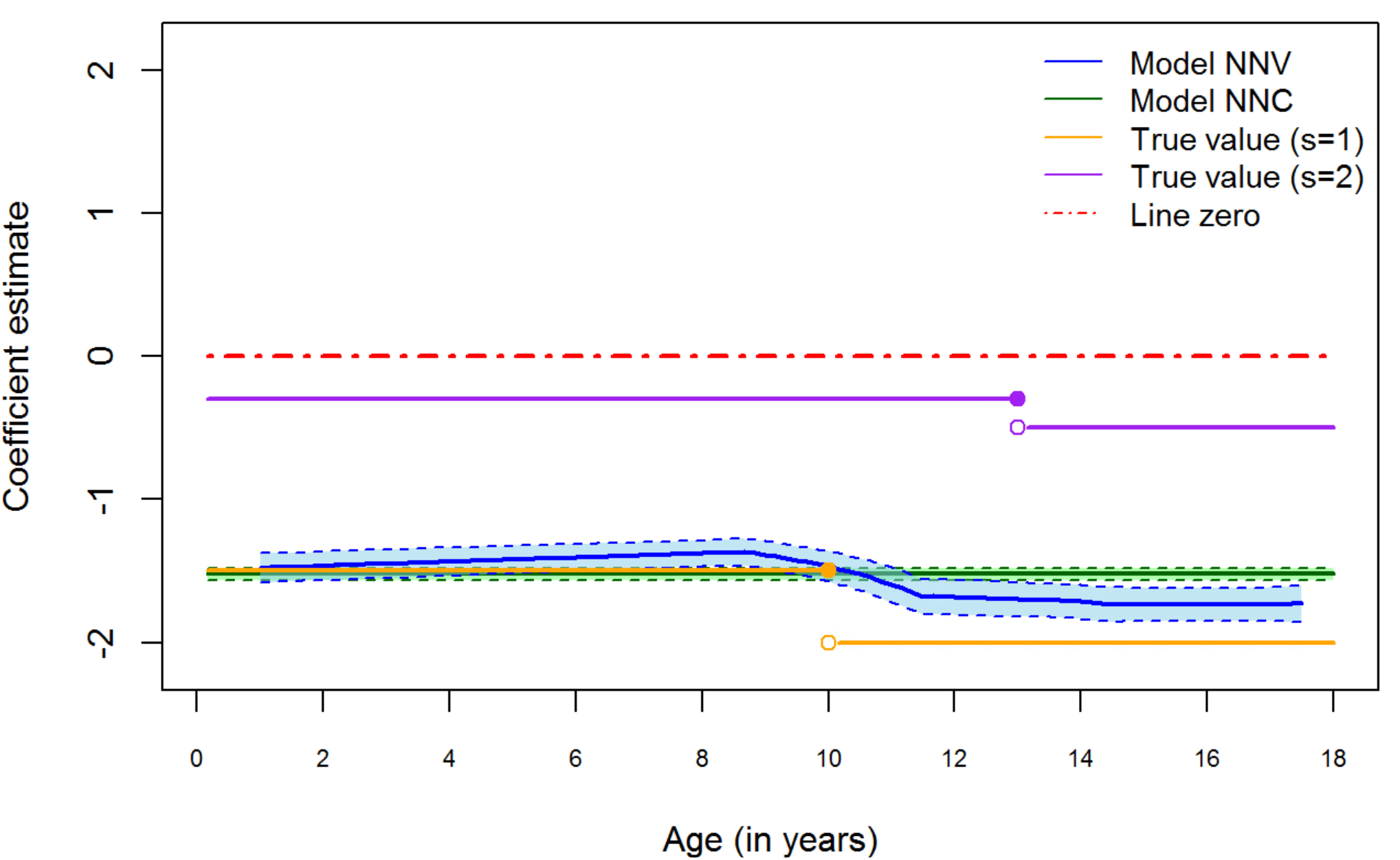}
\caption{Under Model (NNV)} 
\end{subfigure}\hspace*{\fill}
\begin{subfigure}{0.48\textwidth}
\includegraphics[width=\linewidth]{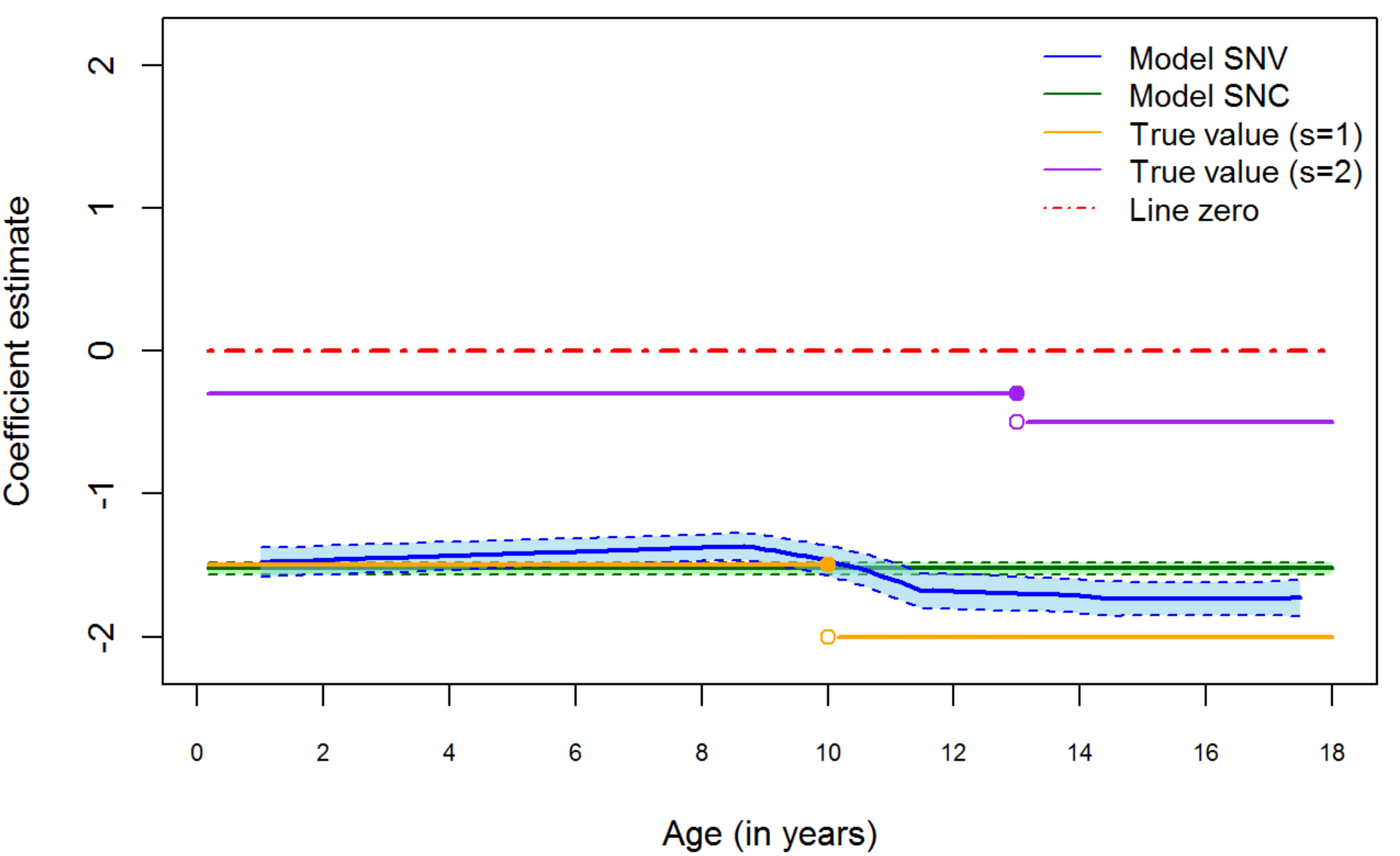}
\caption{Under Model (SNV)} 
\end{subfigure}

\medskip
\begin{subfigure}{0.48\textwidth}
\includegraphics[width=\linewidth]{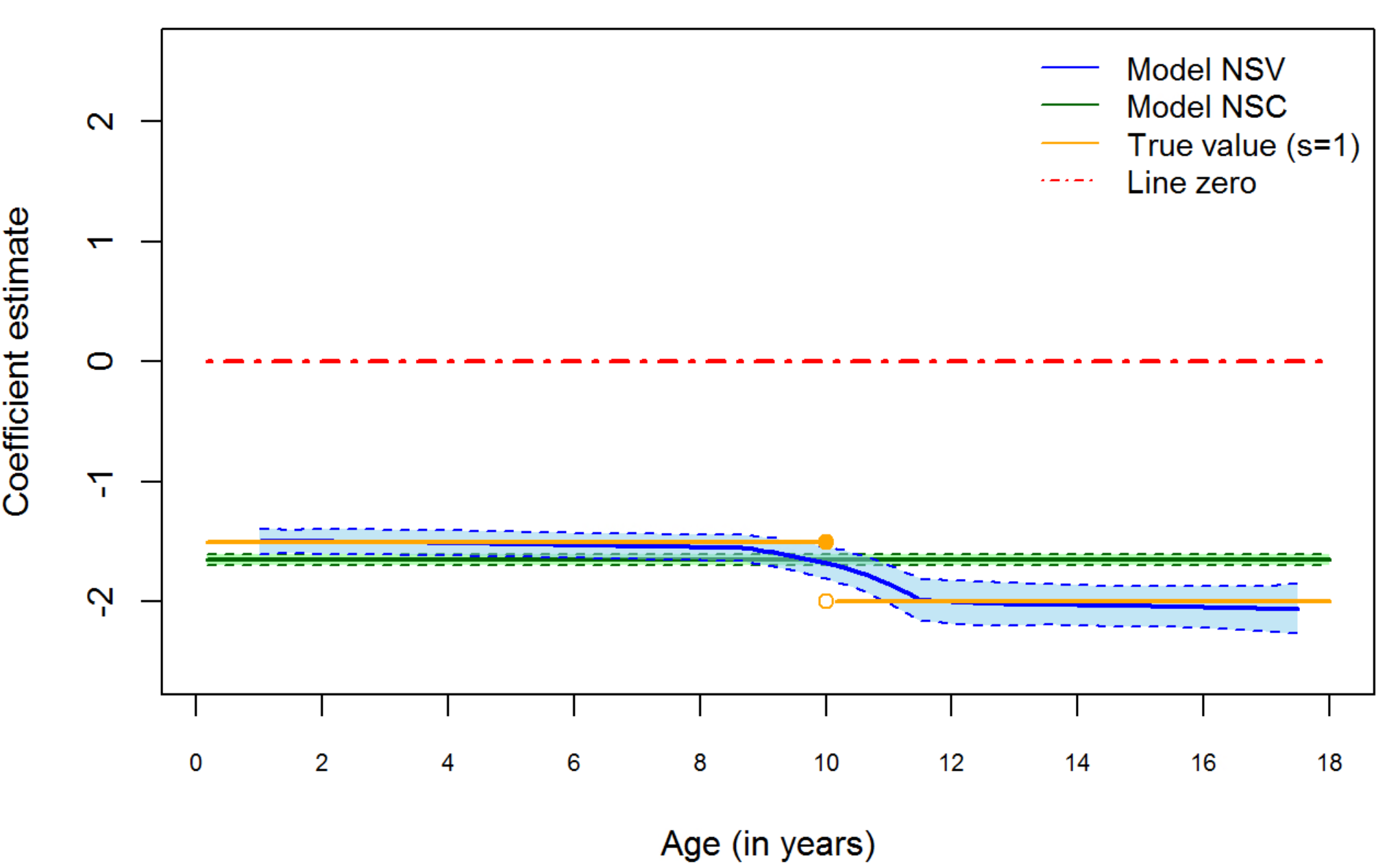}
\caption{Under Model (NSV); stratum 1} 
\end{subfigure}\hspace*{\fill}
\begin{subfigure}{0.48\textwidth}
\includegraphics[width=\linewidth]{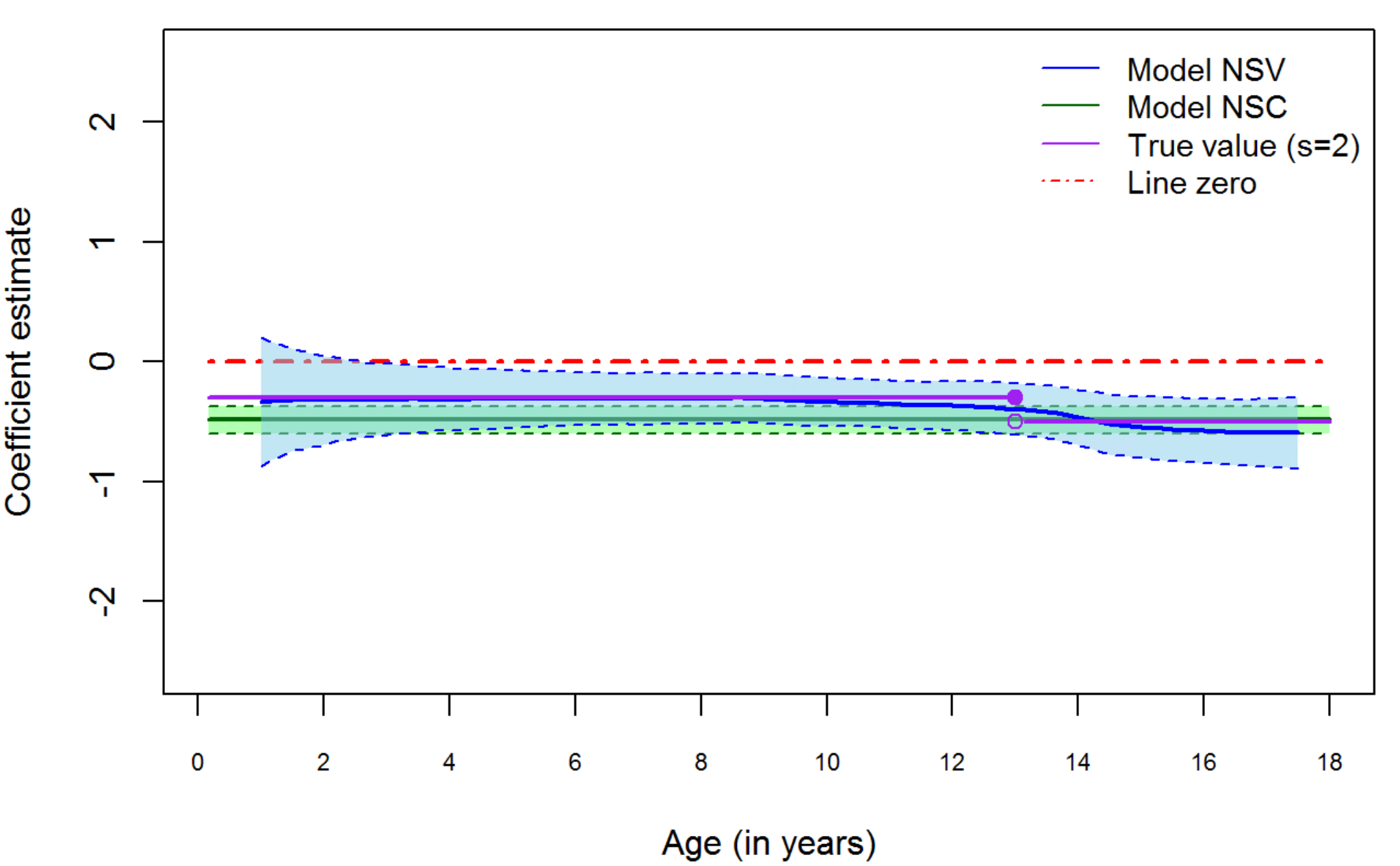}
\caption{Under Model (NSV); stratum 2} 
\end{subfigure}

\medskip
\begin{subfigure}{0.48\textwidth}
\includegraphics[width=\linewidth]{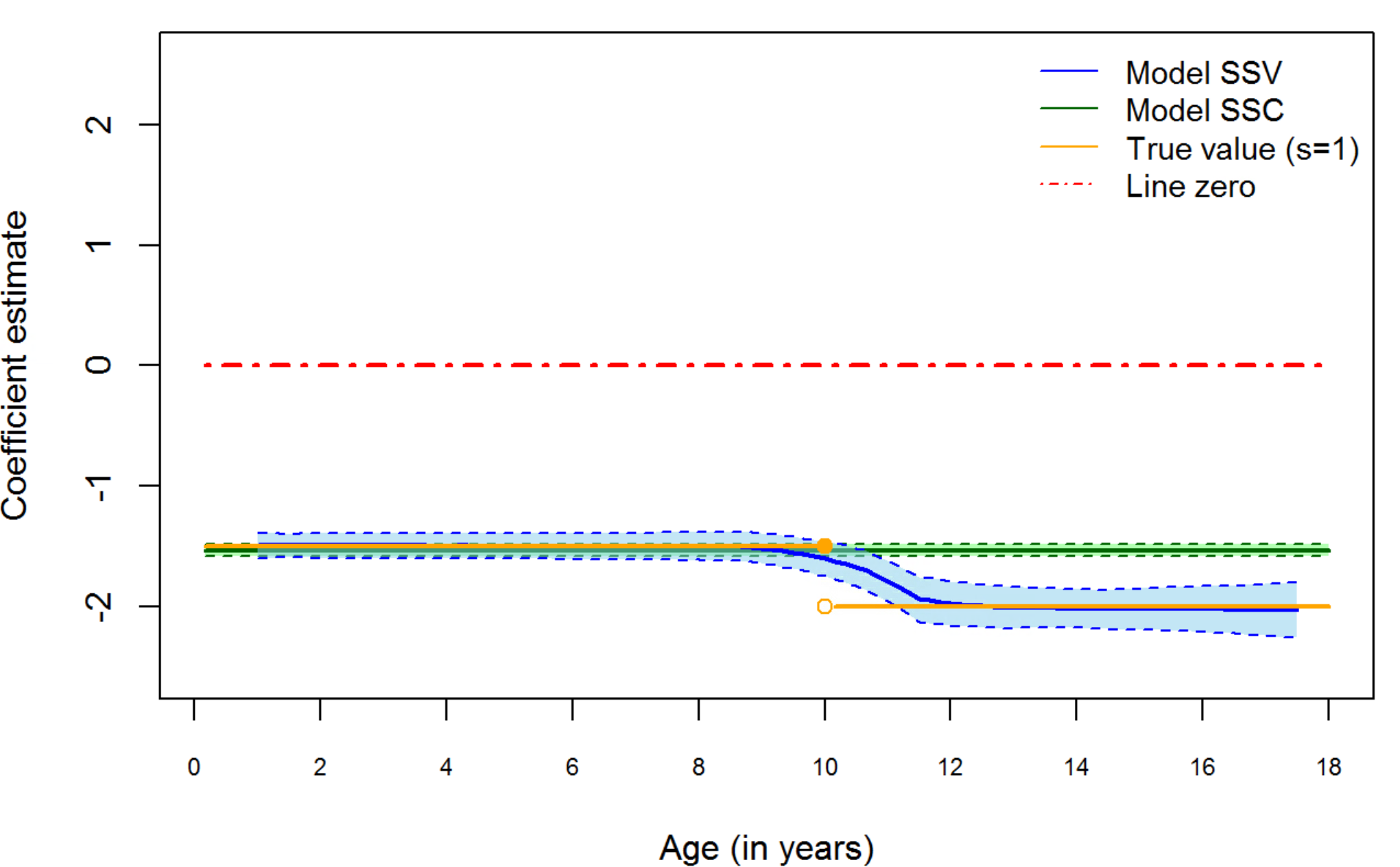}
\caption{Under Model (SSV); stratum 1} 
\end{subfigure}\hspace*{\fill}
\begin{subfigure}{0.48\textwidth}
\includegraphics[width=\linewidth]{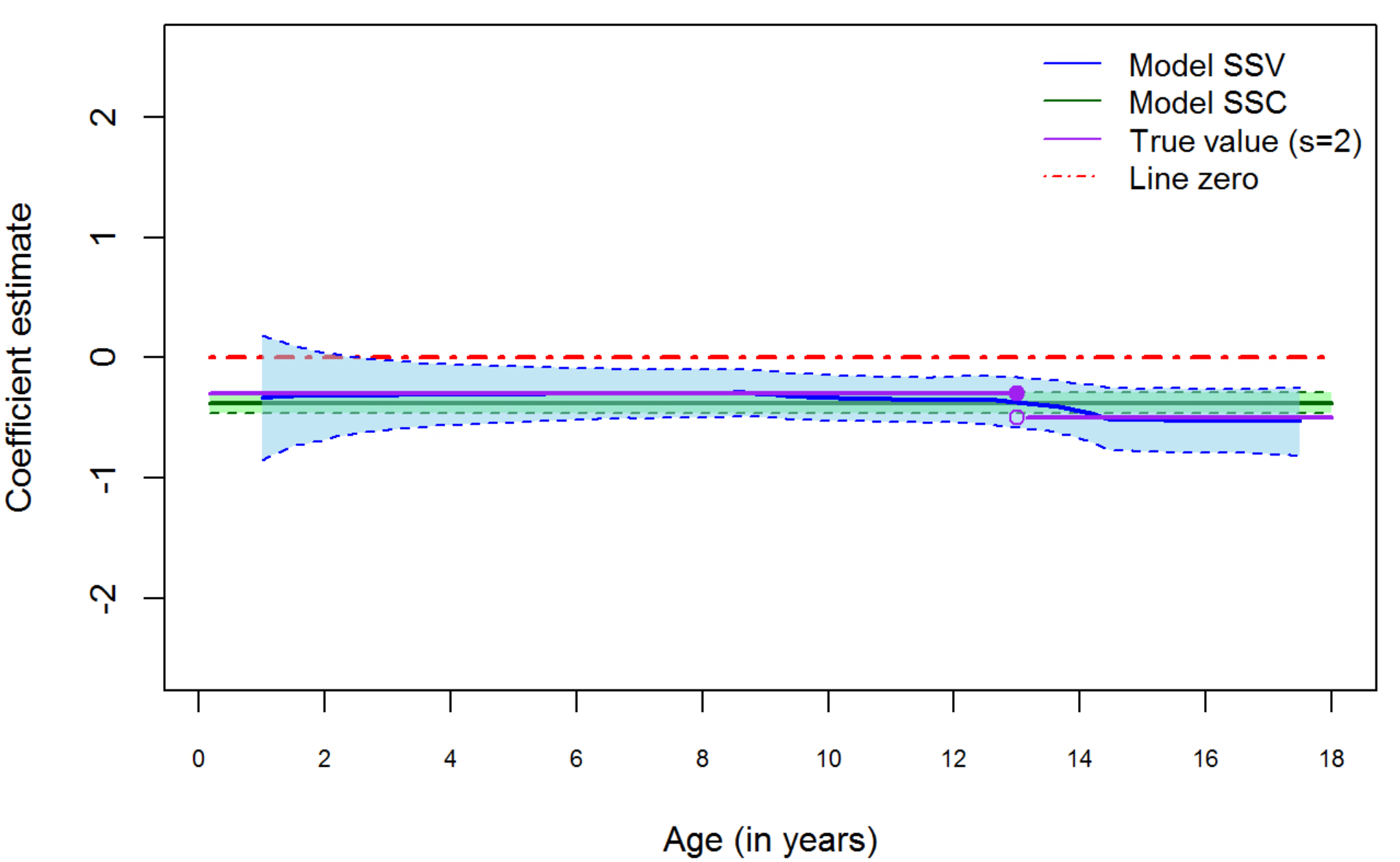}
\caption{Under Model (SSV); stratum 2} 
\end{subfigure}

\caption{Estimated coefficients to the indicator $Z_3$ with 95\% pointwise confidence intervals in Scenario 2 using a 7-year data extraction window based on $1,000$ simulation repetitions.\\ \footnotesize{\noindent Note: The models are shown in Table (\ref{Tab:submodels}).}} \label{fig:S_coef_z3}
\end{figure}

\begin{figure}[t!] 
\begin{subfigure}{0.48\textwidth}
\includegraphics[width=\linewidth]{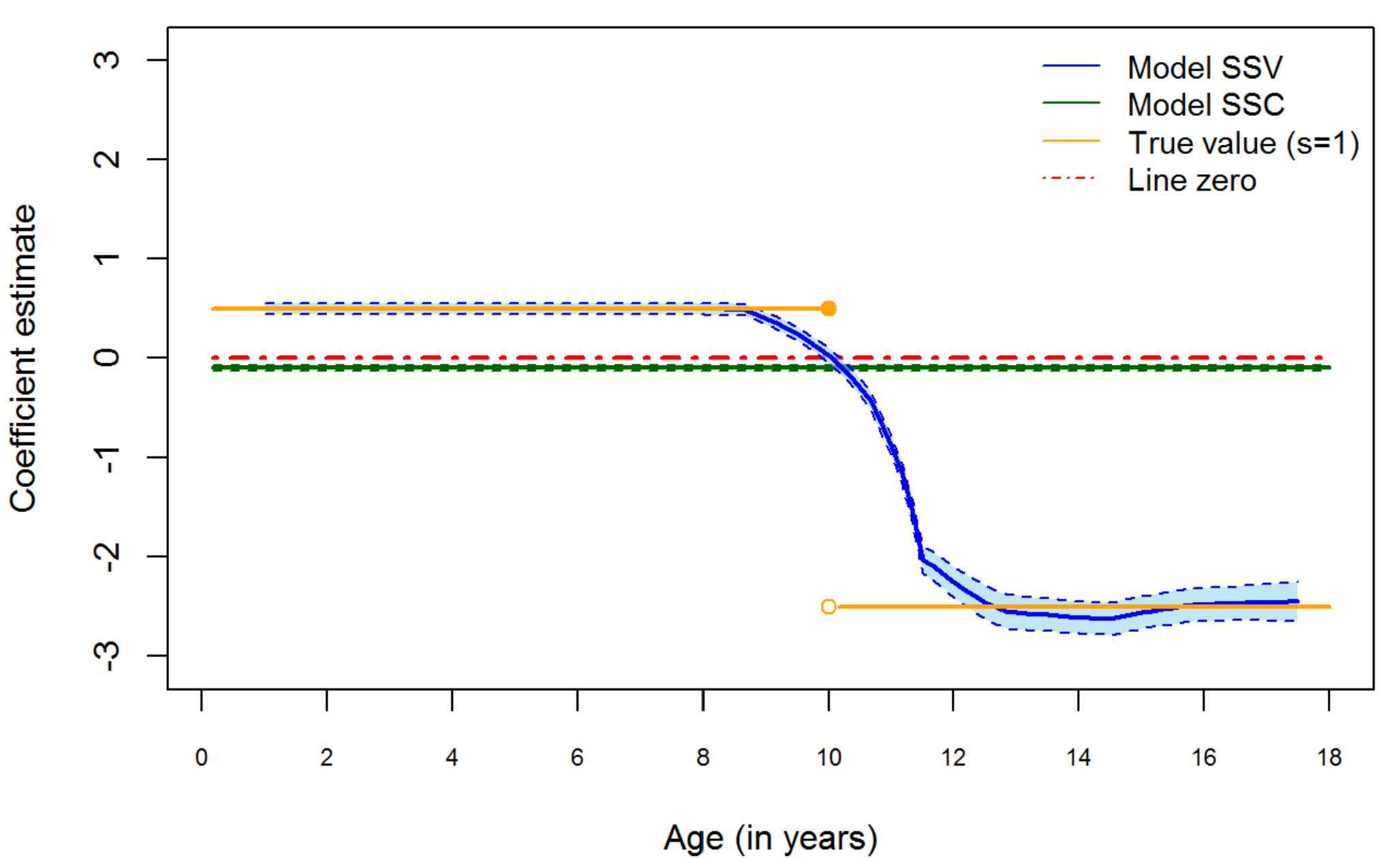}
\caption{Coefficient of $Z_1$ (stratum 1)} 
\end{subfigure}\hspace*{\fill}
\begin{subfigure}{0.48\textwidth}
\includegraphics[width=\linewidth]{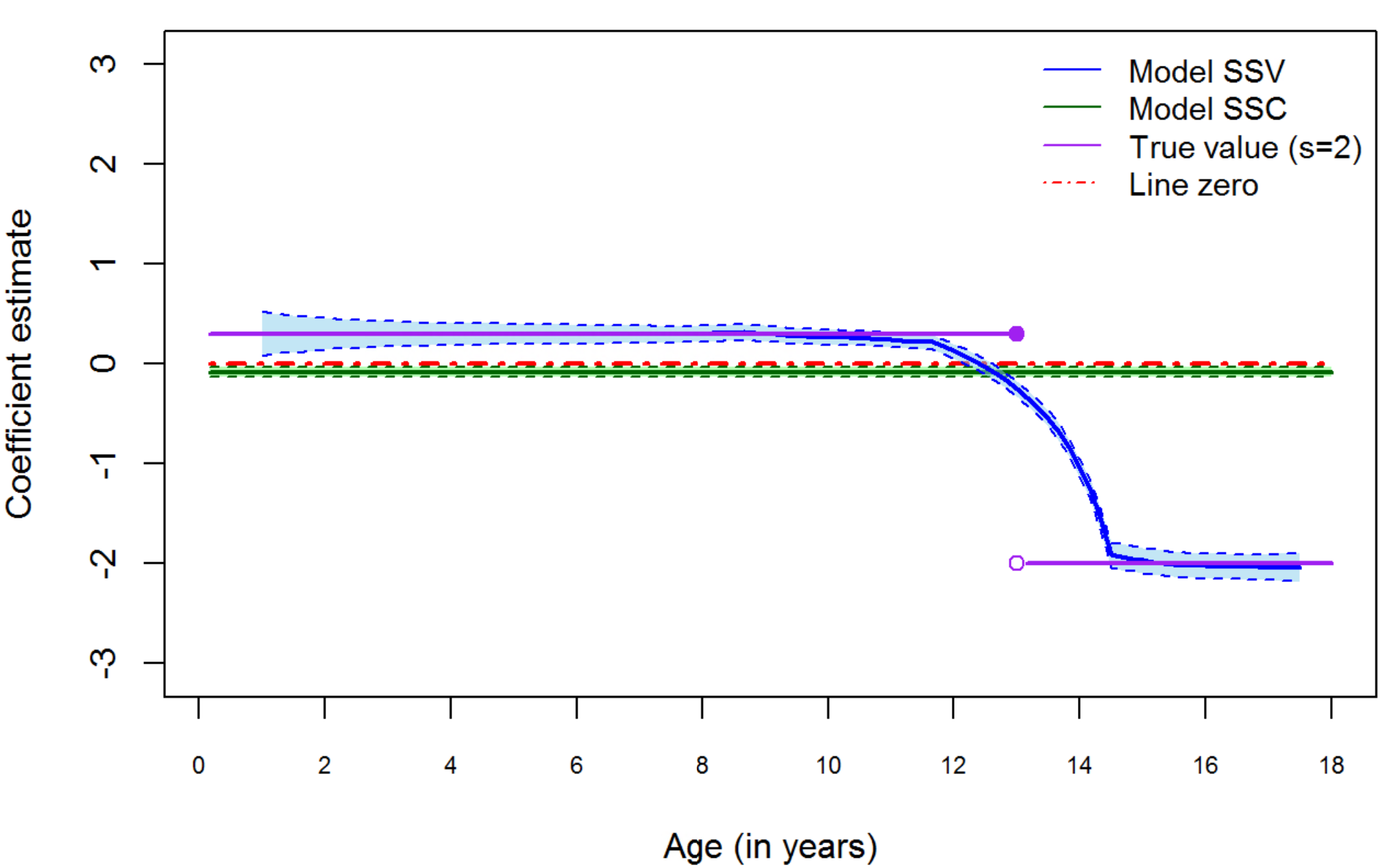}
\caption{Coefficient of $Z_1$ (stratum 2)} 
\end{subfigure}
\caption{Estimated coefficients to the indicator $Z_1$ under Model (SSV) with 95\% pointwise confidence intervals in Scenario 2 using an 18-year data extraction window based on $1,000$ simulation repetitions.\\ \footnotesize{\noindent Note: Model (SSV): $\lambda(a\mid \mathcal{H}_i(a), Z_i)=\lambda_{0s}(a) \exp\{\beta_s(a)^{'} Z_i\}$; Model (SSC): $\lambda(a\mid \mathcal{H}_i(a), $ $Z_i)=\lambda_{0s}(a) \exp\{\beta_s' Z_i\}$.}}\label{fig:paper2_sim_sce2_18yr_SSVV}
\end{figure}

\begin{figure}[t!] 
\centering

\begin{subfigure}{0.5\textwidth}
\includegraphics[width=\linewidth]{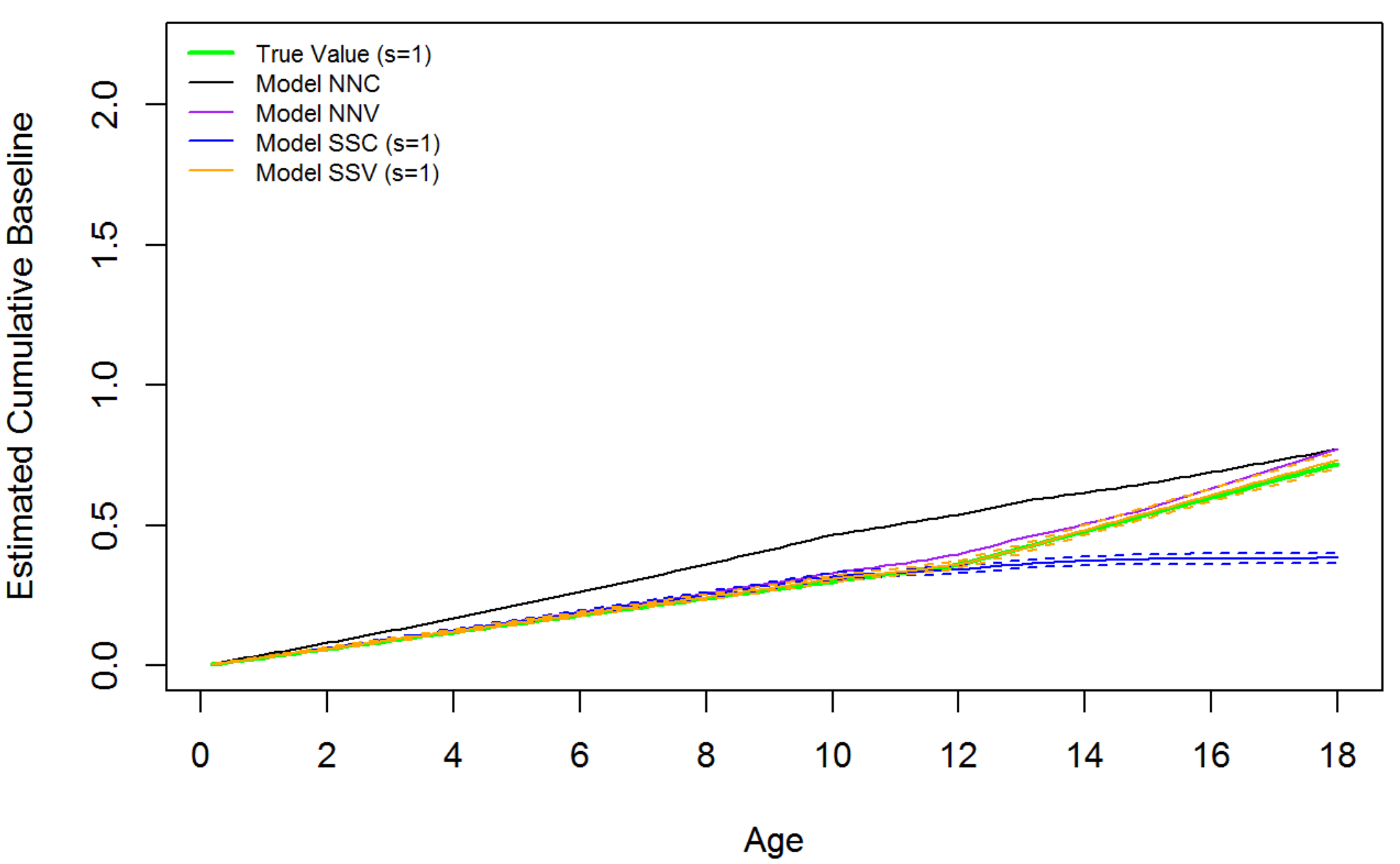}
\caption{Under Model (SSV); stratum 1} 
\end{subfigure}\hspace*{\fill}
\begin{subfigure}{0.5\textwidth}
\includegraphics[width=\linewidth]{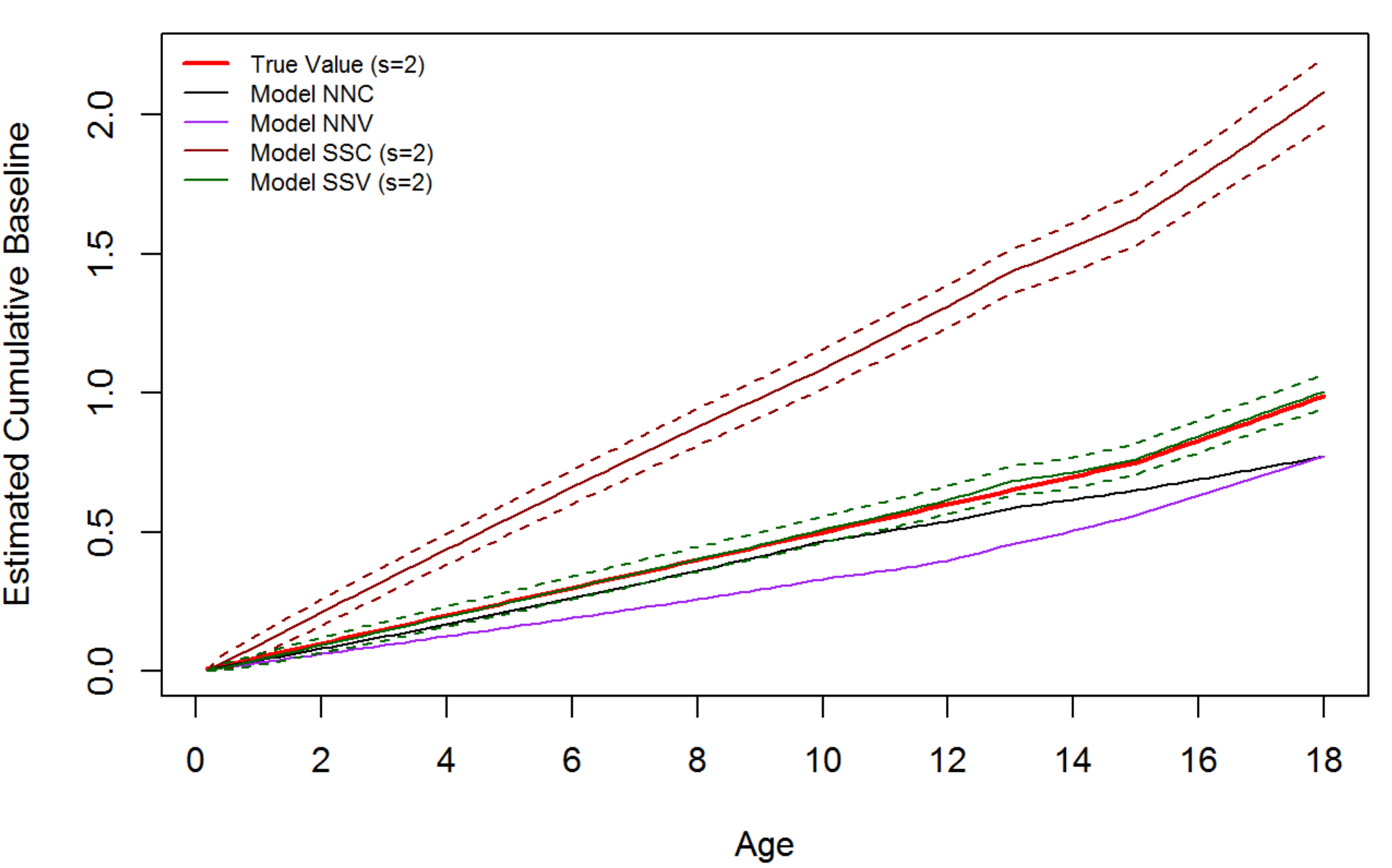}
\caption{Under Model (SSV); stratum 2} 
\end{subfigure}
\caption{\label{Figure:paper2_sim_sce2_cumB}Estimated cumulative baseline intensity functions under Model (SSV) with 95\% pointwise confidence intervals in Scenario 2 using a 7-year data extraction window based on $1,000$ simulation repetitions.\\
\footnotesize{\noindent Note: The models are shown in Table (\ref{Tab:submodels}); Models (NNV) and (NSV) do not incorporate stratification in the baseline intensity functions; to improve readability, we display confidence intervals only for models with time-varying coefficients.}}. 
\end{figure}

\noindent Figures \ref{fig:S_coef}-\ref{fig:S_coef_z3} present the estimated coefficients to the indicators $Z_1$, $Z_2$, and $Z_3$ under various models, with 95\% pointwise confidence intervals. We use Figure \ref{fig:S_coef} as an example to illustrate the estimation results. The orange and purple solid lines represent the true coefficient functions for strata $1$ and $2$, respectively. The blue solid curves denote the time-varying coefficient estimates, and the green solid lines indicate the time-independent estimates. In Scenario 2, where the true model includes stratification in both coefficients and baselines, the first two rows illustrate the estimates under model mis-specification. In Figures (\ref{fig:S_a}) and (\ref{fig:S_b}), the estimated non-stratified coefficients align with the true function of stratum 1 before age $9$ and fall between the two true functions after age $14.5$. This pattern is expected since most subjects belong to stratum 1 at younger ages, and the non-stratified estimates can be considered as a weighted average of the estimates across both strata. Figures (\ref{fig:S_c}) and (\ref{fig:S_d}) show that without stratification in the baselines, the estimated coefficients in both strata catch the true functions well at younger ages but not as well at later ages, especially for the estimates in stratum 1. Figures (\ref{fig:S_e}) and (\ref{fig:S_f}) present that the true functions fall within the 95\% pointwise confidence intervals for most of the time, which verifies the consistency of our proposed estimator. A comparison between Figures (\ref{fig:S_c})-(\ref{fig:S_d}) and (\ref{fig:S_e})-(\ref{fig:S_f}) suggests that the estimates from Model (NSV) are robust even when the true model is (SSV). Furthermore, the estimation results under Model (SSV) improve with a larger data extraction window, as illustrated in Figure \ref{fig:paper2_sim_sce2_18yr_SSVV}. Figure \ref{Figure:paper2_sim_sce2_cumB} presents the estimated cumulative baseline intensity functions under selected models based on a 7-year data extraction window. The green and red solid lines represent the true functions for strata $1$ and $2$, respectively. Among all the models, only the estimates from the proposed Model (SSV) closely align with the true cumulative baseline functions in both strata.

\section{Conclusion}\label{sec:conclusion}

We make three key contributions in this paper. First, we introduce an innovative intensity-based model with stratification, allowing a simple summary of event history to fully specify a counting process. Compared to \cite{Chen&Hu&Rosychuk2025}, our proposed model with time-varying coefficients is more flexible and has more applications, such as dynamic prediction. Specifically, the fitted model can be used to predict the risk of having a future event for a subject at any age $a\in(0,A^\star)$, given covariates and history information \citep{VANHOUWELINGENHANSC.2007DPbL,LerouxAndrew2018Dpif}. Moreover, our proposed model includes the models of \cite{Hu_Rosychuk2016} and \cite{Yi_Hu_Rosychuk2020} as special cases. These studies focused on marginal models, with \cite{Hu_Rosychuk2016} emphasizing inference for the subpopulation $\mathcal{P}_1$ and \cite{Yi_Hu_Rosychuk2020} investigating the evolution of the event patterns across time periods. Second, we propose an approach for estimating time-varying coefficients for zero-truncated recurrent event data integrated with population census information under the stratified model. Third, our findings provide practical insights into how past MHED visits influenced the occurrence of later visits over age. Additionally, with the estimation results, we can estimate the risk of a visit for a certain group, such as Edmonton boys, at a given age with a known event history. 

Some limitations and considerations regarding our proposed method need to be further discussed. First, the approach of combining zero-truncated data with population census information is inherently restricted by the availability and scope of such census data. For example, if covariates like prior mental health conditions are of interest, we need to consider other sources of supplementary information. Additionally, our estimation procedure can be implemented either by specifying the conditional probability $P(Y_i^{(s)}(u)=1|\mathcal{Q}_{1i})$ under the proposed model or by assuming a distribution for the stratification variable $S_i(a)$. Moreover, the local constant estimation method requires a sufficient number of events within each neighbourhood defined by the bandwidth. When data are sparse, the estimation procedure may fail to converge. To address this issue, we recommend adjusting the bandwidth according to the characteristics of the dataset. If needed, a time-varying bandwidth can be employed, though additional adjustments to the variance estimation may be required.

There are several directions for future investigations. In this paper, we assume that birthdates are independent of the counting process. However, this assumption may not hold since different generations could have distinct event patterns. In a future study, we are going to explore the differences in the event patterns over time. For instance, we can study the MHED visit patterns before, during, and after the COVID-19 pandemic. Additionally, we plan to account for potential correlations among subjects within the same community (or other small regions) and extend the analysis to different types of events, such as MHED visits of different severity levels.

\appendix                       
\bigskip

\section{Asymptotic Proofs for Proposition 1 in Section \ref{sec:estimation_P_avail_S}}
\label{sec:Appendix_proofs} 
\small
We adapted the conditions in \cite{Hu_Rosychuk2016} for establishing the consistency and asymptotic normality of the coefficient estimators derived in Section \ref{sec:estimation_P_avail_S} under Model (\ref{eq:model}).
\begin{enumerate}[I.]
    \item $\{N_i(\cdot),Z_i,B_i\}$ for $i=1,\cdots,n=|\mathcal{O}|$ are independent and identically distributed;
    \item $P(C_{L}<\tau_L)>0$ and $P(C_{R}>\tau_R)>0$ for $0<\tau_L<\tau_R<A^\star$, where $\tau_L$ and $\tau_R$ are predetermined constants; 
    \item $N(A^{\star})$ is bounded by a constant with probability 1;
    \item $\lambda_{0s}(a)>0$ and is continuous and all the components of $\beta_s(a)$ have a continuous second derivative for $a\in (0,A^\star)$ and $s\in\mathcal{S}$;
    \item $S(\cdot)$ is a left-continuous step function over $(0, A^\star)$; 
    \item  The kernel function $K_h(\cdot)$ is a bounded and symmetric density with a bounded support;
    \item As $\triangle l\xrightarrow{}0$, $\widehat{P}(Y^{(c)}(a)=1,Z=z) \xrightarrow{unif.}P(Y^{(c)}(a)=1,Z=z)$ for all $a \in (0,A^\star)$ and $z\in \mathcal{Z}$, where $\widehat{P}(Y^{(c)}(a)=1,Z=z)=\sum_{l}\mathcal{C}(l,z, a)/n$ is the estimated probability based on some supplementary information and $l$ denotes how often the information is recorded.

\end{enumerate}

We provide an outline of the proof for Proposition 1 in Section \ref{sec:estimation_P_avail_S}, as detailed below. 
When the stratification is partially known, we consider the following estimating functions: for a fixed $a\in[\tau_L,\tau_R]$ and $s\in\mathcal{S}$,
\small
 \begin{align*}
\begin{split}
\tilde{U}_{s}(\boldsymbol{\gamma};a|\boldsymbol{\lambda}_{0}(\cdot))
&= \sum_{i\in \mathcal{O}_1}^{}\int_{0}^{A^\star} K_h(u-a) P\bigl(Y_i^{(s)}(u)=1|\mathcal{Q}_{1i}\bigr)\big\{Z_i\\
&\hspace{4.5cm}-\tilde{\bar{Z}}_s(\boldsymbol{\gamma};u)\big\} Y_i^{(c)}(u) dN_i(u),
    \end{split}
    \end{align*}
\normalsize
and
\small
\begin{align*}
\begin{split}
    \tilde{V}_{s}(\boldsymbol{\lambda}_{0}(\cdot);a\big|\boldsymbol{\gamma})&=\sum_{i\in \mathcal{O}_1}^{}P\bigl(Y_i^{(s)}(a)=1|\mathcal{Q}_{1i}\bigr)Y_i^{(c)}(a)dN_i(a)\\
    &\hspace{0.2cm}
    -\sum_{i\in\mathcal{O}}^{}\hat{E}\big[Y_i^{(s)}(a)Y_i^{(c)}(a)\exp\{\gamma_s^{'}Z_i\}\big]\lambda_{0s}(a)da,
\end{split}
\end{align*}
where $\tilde{\bar{Z}}_s(\boldsymbol{\gamma};u)=\sum_{i\in\mathcal{O}}^{}\hat{E}\big[Y_i^{(s)}(u)Y_i^{(c)}(u)Z_i\exp\{\gamma_s^{'}Z_i\}\big]/\sum_{i\in\mathcal{O}}^{}\hat{E}\big[Y_i^{(s)}(u)$ $Y_i^{(c)}(u)$ $\exp\{\gamma_s^{'}Z_i\}\big]$
with $\sum_{i\in\mathcal{O}}^{}\hat{E}\big[Y_i^{(s)}(u)$ $Y_i^{(c)}(u)Z_i^{\otimes q}\exp\{\gamma_s^{'}Z_i\}\big]=\sum_{z\in \mathcal{Z}}^{}$ $\Bigl\{
P\bigl(Y^{(s)}(u)=1|Z=z\bigr)z^{\otimes q}\exp\{\gamma_s^{'}z\}\bigl[\sum_{l}$ $\mathcal{C}(l,z,\lfloor u \rfloor)\bigr]\Bigr\}$ for $q=0, 1, 2$. Let $EF_n\big(\boldsymbol{\theta}(\cdot);a\big)=1/n\Big(\big(\tilde{U}_{1}(\boldsymbol{\gamma};a|\boldsymbol{\lambda}_{0}(\cdot))', \tilde{V}_{1}(\boldsymbol{\lambda}_{0}(\cdot);a\big|\boldsymbol{\gamma})\big)',\cdots, \big(\tilde{U}_{S^\star}(\boldsymbol{\gamma};a|\boldsymbol{\lambda}_{0}(\cdot))'$ $,\tilde{V}_{S^\star}(\boldsymbol{\lambda}_{0}(\cdot);a\big|\boldsymbol{\gamma})\big)'\Big)'$, where $\mathcal{S}=(1,\cdots, S^\star)$ and $\boldsymbol{\theta}(a)=\bigr(\boldsymbol{\gamma}^{(a)'},\boldsymbol{\lambda}_{0}(a)^{'}\bigl)^{'}$ for $a\in(0,A^\star)$. We introduce the hyperindex to emphasize that the parameter $\boldsymbol{\gamma}$ is defined specifically with respect to age $a$. Let 
$\boldsymbol{\theta}_0(a)=\bigr(\boldsymbol{\gamma}_0^{(a)'},\boldsymbol{\lambda}_{00}(a)^{'}\bigl)^{'}$ for $a\in(0,A^\star)$ be the true value of $\boldsymbol{\theta}(a)$ and $\tilde{\boldsymbol{\theta}}(a)=\bigr(\tilde{\boldsymbol{\gamma}}^{(a)'},\tilde{\boldsymbol{\lambda}}_{0}(a)^{'}\bigl)^{'}$ for $a\in(0,A^\star)$ be our proposed estimators.  By the Taylor expansion of $EF_n\big(\tilde{\boldsymbol{\theta}}(\cdot);a\big)$ about $\boldsymbol{\theta}_0(a)$ for $\tilde{\boldsymbol{\theta}}(a)$ in the neighborhood of $\boldsymbol{\theta}_0(a)$, we have
\begin{align*}
\begin{split}
   EF_n\big(\tilde{\boldsymbol{\theta}}(\cdot);a\big)&= EF_n\big(\boldsymbol{\theta}_0(\cdot);a\big)+ \frac{\partial EF_n\big(\boldsymbol{\theta}_0(\cdot);a\big)}{\partial \boldsymbol{\theta}}(\widetilde{\boldsymbol{\theta}}-\boldsymbol{\theta}_0)+o\big((\widetilde{\boldsymbol{\theta}}-\boldsymbol{\theta}_0)\big).
\end{split}
\end{align*}
\normalsize
\noindent By Lemma 1, we can establish the consistency of $\widetilde{\boldsymbol{\theta}}(\cdot)$, which includes $\widetilde{\boldsymbol{\beta}}(\cdot)$ as a component. Building on Lemmas 1 and 2 and the consistency of $\widetilde{\boldsymbol{\theta}}(\cdot)$, we can further derive the asymptotic normality of $\widetilde{\boldsymbol{\theta}}(\cdot)$, thereby completing the proof of Proposition 1.

\begin{lemma}
    As $n\rightarrow \infty$, $\sup_{a\in(0,A^\star)}|EF_n\big(\boldsymbol{\theta}_0(\cdot);a\big)|\xrightarrow{a.s.}\boldsymbol{0}$ and $\partial EF_n\big(\boldsymbol{\theta}_0(\cdot);$ $a\big)/{\partial \boldsymbol{\theta}}\xrightarrow{a.s.}\Pi\big(\boldsymbol{\theta}_0(\cdot);a\big)$ for $a\in(0,A^\star)$, where $\Pi\big(\boldsymbol{\theta}_0(\cdot);a\big)$ is a positive-definite matrix. 
\end{lemma}

\begin{lemma}
    As $n\rightarrow \infty$, $\sqrt{nh}EF_n\big(\boldsymbol{\theta}_0(\cdot);a\big)\xrightarrow{w.}\mathcal{G}(a)$ with the bandwidth $h=O(n^{-v})$ for $1/2<v<1$ and $a\in(0,A^\star)$, where $\mathcal{G}(a)$ is a Gaussian process with mean zero. 
\end{lemma}





\section{Details for Estimation Procedures in Section \ref{sec:estimation}}
\subsection{The calculation of the probability $P\bigl(Y^{(s)}(a)=1|Z=z\bigr)$ with the stratification variable (\ref{eq:stratification_variable})}\label{subsec:Append_P_s_Z}
Under the stratification variable (\ref{eq:stratification_variable}), the probability is
\small
\begin{align*}
    P(Y^{(s)}(a)=1|Z=z)&=P(S(a)=s|Z=z)\\
    &=\begin{cases}exp\Big\{-\int_{0}^{a} \lambda_{01}(u) \exp\{\beta_1(u)' z\}du\Big\} &\text{ }s=1 \\1-exp\Big\{-\int_{0}^{a} \lambda_{01}(u) \exp\{\beta_1(u)' z\}du\Big\} &\text{ }s=2\end{cases}.
\end{align*}
\bigskip

\subsection{The calculation of the probability $P\bigl(Y_i^{(s)}(a)=1|\mathcal{Q}_{1i}\bigr)$ with the stratification variable (\ref{eq:stratification_variable})}{\label{subsec:Append_p_s_data}}
\small
For $i\in \mathcal{O}_1$ and $a\in(C_{L_i},C_{R_i}]$, the probability
\begin{align*}
    P\bigl(Y_i^{(1)}(a)=1|\mathcal{Q}_{1i}\bigr)
    &=P(N_i(a-)=0|\mathcal{Q}_{1i})\\
    &=\begin{cases}0 &\hspace{-2.3cm}\exists\text{ } a_{ij}\in(C_{L_i},C_{R_i}],~a_{ij}<a\\P\big(N_i(C_{L_i})=0,N_i(a-)-N_i(C_{L_i})=0|\mathcal{Q}_{1i}\big) &\text{otherwise}\end{cases},
\end{align*}
where $a_{ij}$ is the age of subject $i$ at the $j$-th observed event time within $(C_{L_i},C_{R_i}]$ for $j=1,\cdots,N_i^{\star}$ with $N_i^{\star}=N_i(C_{R_i})-N_i(C_{L_i})$. 

After some simplification,
\begin{align*}
    &P\big(N_i(C_{L_i})=0,N_i(a-)-N_i(C_{L_i})=0|\mathcal{Q}_{1i}\big)\\
    &=\frac{1}{Dr}\Bigl[\lambda_{01}(a_{i1})e^{\beta_1(a_{i1})^{'}Z_i}\exp\{-\int_{0}^{a_{i1}}\lambda_{01}(u)e^{\beta_1(u)^{'}Z_i}du\}\Bigr],
\end{align*}
where $Dr=\lambda_{01}(a_{i1})e^{\beta_1(a_{i1})^{'}Z_i}\exp\{-\int_{0}^{a_{i1}}\lambda_{01}(u)e^{\beta_1(u)^{'}Z_i}du\}
+\lambda_{02}(a_{i1})e^{\beta_2(a_{i1})^{'}Z_i}$
$\bigg[1-\exp\{-\int_{0}^{C_{L_i}}\lambda_{01}(u)e^{\beta_1(u)^{'}Z_i}du\}\bigg]\\
\times\exp\{-\int_{C_{L_i}}^{a_{i1}}\lambda_{02}(u)e^{\beta_2(u)^{'}Z_i}du\}$.\\


\section*{Competing interests}
No competing interest is declared.


\section*{Acknowledgements}
An operating grant from the Canadian Institutes of Health Research (CIHR) supported
data extraction. This work was supported by individual Discovery Grants from the Natural
Sciences and Engineering Research Council of Canada (NSERC) held by L Zeng, XJ Hu and RJ
Rosychuk, an NSERC Discovery Accelerator Supplement held by XJ Hu, and a Graduate Dean Entrance Scholarship held by AA Chen. 
\bigskip

\noindent \textit{Disclaimer}:
This study is based on data provided by Alberta Health. The interpretations and conclusions are solely those of the authors and do not reflect the views of Alberta Health or the Government of Alberta.

\newpage
\bibliography{myref}

\label{lastpage}

\end{document}